\documentclass[final,6p,times,twocolumn,floatfix,authoryear,nofootinbib,eqsecnum]{revtex4-2}

\usepackage{bm}
\usepackage{newtxmath,newtxtext}

\usepackage{graphicx}
\usepackage{adjustbox}
\usepackage{textcomp}
\usepackage{float}
\usepackage{booktabs}
\usepackage{dcolumn}
\usepackage{ragged2e}
\usepackage{hyperref}

\hypersetup{colorlinks=true,citecolor=blue, linkcolor=blue,filecolor=blue,    urlcolor=blue}

\usepackage{xcolor}
\usepackage{epsfig}
\usepackage{caption}
\usepackage{appendix}
\usepackage{subcaption}

\usepackage{orcidlink}
\usepackage{epsfig}
\usepackage{caption}

\usepackage{float}

\usepackage{multirow}
\usepackage{tcolorbox}
\usepackage{orcidlink}

\newcommand{\bq}{\begin{equation}}
	\newcommand{\eq}{\end{equation}}
\newcommand{\bqn}{\begin{eqnarray}}
	\newcommand{\eqn}{\end{eqnarray}}

\newcommand{\pp}{\partial}

\newcommand{\cs}{\ensuremath{c_{\rm s}^{2}}}

\newcommand{\ti}[1]{\ensuremath{\tilde{#1}}}
\newcommand{\nb}{\ensuremath{ \nabla }}

\begin{document}

	\title{Interacting $k$-essence field with non-pressureless Dark Matter: Cosmological Dynamics and Observational Constraints}
	\author{Saddam Hussain \orcidlink{0000-0001-6173-6140}}
	\email{saddamh@zjut.edu.cn}
	
	\author{Qiang Wu}
	\email{Corresponding author: wuq@zjut.edu.cn}
	
	\author{Tao Zhu}\email{Corresponding author: zhut05@zjut.edu.cn} 
	
	\affiliation{Institute for Theoretical Physics and Cosmology, Zhejiang University of Technology, Hangzhou 310023, China}

	\begin{abstract}
		
		We investigate a class of interacting dark energy and dark matter (DM) models, where dark energy is modeled as a $k$-essence scalar field with an inverse-square potential. Two general forms of interaction are considered: one proportional to the Hubble parameter, and another independent of the Hubble parameter, depending instead on combinations of the energy densities and pressures of the dark sectors. {The cosmological evolution is reformulated in terms of an autonomous system of equations, which provides a convenient phase-space parametrization for the numerical integration of the background dynamics and for confronting the models with observations.} The models are tested against a wide range of observational datasets, including cosmic chronometers (CC), BAO measurements from DESI DR2, compressed Planck data (PLA), Pantheon+ (PP), DES supernovae, Big Bang Nucleosynthesis (BBN), and strong lensing data from H0LiCOW (HCW). The analysis shows that the models consistently reproduce all major cosmological epochs and yield statistically competitive results compared to the flat $\Lambda$CDM model. The models exhibit late-time de-Sitter solutions, ensuring ghost-free evolution, with the Hubble constant in the range $H_0 \sim 67$--$70$ km/s/Mpc.
		
	\end{abstract}

	\date{\today}
	\maketitle
	\flushbottom
	
	\section{Introduction}
	
	Since the first observational evidence from Type Ia supernovae \cite{SupernovaCosmologyProject:1998vns,SupernovaSearchTeam:1998fmf}, several independent probes including the Cosmic Microwave Background (CMB), large-scale structure, Baryon Acoustic Oscillations (BAO), and strong and weak gravitational lensing \cite{WMAP:2003elm,Sherwin:2011gv,Wright:2007vr,DES:2016qvw,DES:2021esc,SDSS:2005xqv,Suyu:2013kha,ACTPol:2014pbf,Planck:2015fie,SDSS:1999zww,Sandage:2006cv} have confirmed the existence of a component responsible for the accelerating expansion of the Universe. Observations indicate that the dominant constituents of the cosmic energy budget are dark energy (DE), which drives the accelerated expansion and contributes roughly $70\%$, and dark matter (DM), a gravitating but electromagnetically neutral component that makes up about $26\%$ and plays a crucial role in structure formation \cite{Primack:1997av,DelPopolo:2007dna,Diao:2023tor,Mina:2020eik,Blumenthal:1984bp}. 
	
	The most widely accepted model that successfully explains observations on both large and small scales is the $\Lambda$CDM model, which consists of a cosmological constant $\Lambda$—exerting negative pressure and mimicking an anti-gravitating effect—coupled with a pressureless perfect fluid known as cold dark matter (CDM). The effective equation of state (EoS), i.e., the ratio of pressure to energy density, in this model is $w=-1$. Although $\Lambda$CDM is remarkably successful observationally, it faces significant theoretical challenges, including the coincidence problem and the discrepancy between the theoretically predicted and observationally inferred values of the cosmological constant, which differ by about 120 orders of magnitude \cite{Copeland:2006wr,Weinberg:1988cp,Rugh:2000ji,Padmanabhan:2002ji,Carroll:1991mt}.

	Recent measurements of the Hubble constant \(H_0\) have sparked intense debate regarding the validity of the standard cosmological model. The most recent CMB measurements yield a value of $67.4 \pm 0.5$ km/s/Mpc \cite{Planck:2018vyg}, whereas the SH0ES team, using the distance-ladder approach calibrated with Cepheid variables, reports $73.2 \pm 1.3$ km/s/Mpc \cite{Riess:2020fzl}. This discrepancy exceeds the $4\sigma$ level. Several other independent probes, both direct and indirect, also report tensions at more than the $4\sigma$ level \cite{Freedman:2020dne,H0LiCOW:2019pvv,Pesce:2020xfe,SPT-3G:2021eoc,Cuceu:2019for,Pascale:2024qjr}. Furthermore, the recent release of BAO observations from the Dark Energy Spectroscopic Instrument (DESI R1 and R2) has drawn significant attention, as the results challenge the assumption of a constant dark energy equation of state $(w=-1)$ at more than the $3\sigma$ level \cite{DESI:2024mwx,Cortes:2024lgw,DESI:2025fii,Hussain:2025nqy,Arora:2025msq,Scherer:2025esj,DESI:2025zgx,Huang:2025som,Li:2024qso}. These results suggest a preference for a dynamical form of dark energy over the cosmological constant $\Lambda$.
	
	The accumulated shortcomings of the \(\Lambda\)CDM model have motivated the exploration of many alternatives to describe dark energy, including scalar fields, modified gravity, interacting dark energy, and unified dark sector models \cite{Wang_2024,Lima:2004cq,Wu:2025wyk,Clifton:2011jh,Gomez-Valent:2023hov,Giani:2024nnv,Zimdahl:2001ar,Bertacca:2010ct,Ansoldi:2012pi,Yao:2024kex,Gross:1986mw,Bento:1995qc,Nojiri:2005vv,Tsujikawa:2006ph,Hussain:2024yee,Hussain:2025vbo,Odintsov:2019clh}. Among these, potential-driven quintessence scalar fields have gained popularity due to their simple formulation and ability to closely mimic the fiducial model under suitable conditions \cite{Peebles:2002gy,Nishioka:1992sg,Ferreira:1997hj,Copeland:1997et}. However, these models often suffer from fine-tuning problems associated with potential parameters. To alleviate such issues, kinetically driven scalar fields inspired by string theory--commonly referred to as $k$-essence--have been considered \cite{Armendariz-Picon:2000nqq,Armendariz-Picon:2000ulo,Chiba:1999ka,Fang:2014qga,Armendariz-Picon:1999hyi,Armendariz-Picon:2005oog,Arkani-Hamed:2003pdi,Scherrer:2004au,Chatterjee:2021ijw,Hussain:2022osn,Bhattacharya:2022wzu,Hussain:2022dhp}.

	In particular, the $k$-essence field has been studied with quadratic kinetic forms of \(F(X)\), where \(X = -\frac{1}{2} \nabla_{\mu}\phi \nabla^{\mu}\phi\), in the presence of inverse-square potentials. Such constructions can drive late-time cosmic acceleration and generate distinct cosmological phases when minimally coupled with background fluids such as dark matter and radiation. Due to the nonlinear kinetic term, the field can exhibit both quintessence-like $(w>-1)$ and phantom-like $(w<-1)$ behavior without additional fine-tuning. Moreover, under a constant potential, the field has been shown to unify dark matter and dark energy, making $k$-essence an attractive candidate \cite{Scherrer:2004au}.

	Despite its theoretical appeal, only a few attempts have been made to test $k$-essence against a wide range of observational data \cite{Yang:2009zzl,Dinda:2023mad,Hussain:2024qrd}. A recent study analyzed a well-established $k$-essence Lagrangian, $\mathcal{L}_{\phi} = -V(\phi)(-X+X^2)$, with exponential and inverse-square potentials against late-time data \cite{Hussain:2024qrd}. With minimal coupling to background fluids, the field exhibited stable phantom-like behavior for the considered datasets. Although phantom-like dark energy has appeared in many parameterized models--especially in light of DESI BAO results--phantom fields are generally considered unphysical, as they introduce ghosts and instabilities at the perturbation level \cite{Caldwell:2003vq,Ludwick:2017tox}.
	
	To address these issues, we introduce an additional degree of freedom by allowing an interaction between the $k$-essence field and dark matter, implemented by modifying the conservation equations to permit energy exchange between the components. While such an interaction is phenomenological, it provides an effective mechanism to study non-gravitational couplings between the dark sectors \cite{Zimdahl:2001ar,Wang:2016lxa,Lee:2006za,Bandyopadhyay:2017igc,Kritpetch:2024rgi,Li:2025owk}. Specifically, we construct a general interacting model incorporating terms involving the Hubble parameter, dark matter and dark energy densities $(\rho)$, pressures $(P)$, and the kinetic term $(X)$. We consider two classes of interaction: in the first case, the interaction is proportional to $H$ and other terms, i.e.,
	$Q \propto H \left( \sum_{i} \rho_{i}+ P_{i}\right) X^{\gamma},$
	while in the second case, it is proportional only to the pressure and density terms with a constant Hubble scale,
	$Q \propto H_0 \left( \sum_{i} \rho_{i}+ P_{i}\right) X^{\gamma}.$ 
	These models are referred to as Model A and Model B, respectively. 
	
	{We study the cosmological evolution of these models by reformulating the background equations as an autonomous dynamical system, which provides a convenient phase-space description for the numerical analysis, and we constrain the model parameters using a diverse set of cosmological observations, including cosmic chronometers (CC), DESI BAO (BAO), compressed Planck likelihood (PLA), DES5YSN (DES), Pantheon+ (PP), Big Bang Nucleosynthesis (BBN), and H0LiCOW (HCW) data.} By exploring a wide parameter space and applying observational constraints, we numerically evaluate the cosmological behavior of the models and assess their stability. Furthermore, by incorporating this diverse dataset, spanning from the early to the late Universe, we also probe the nature of dark matter. While dark matter is usually assumed to be pressureless and to follow the same profile as baryonic matter, here we adopt a more general equation of state (EoS) parameter \(w\) for the dark matter fluid and constrain it in light of the combined observations.
	
	We outline the structure of this work as follows. In Sec.~\ref{sec:basic_frame}, we present the action for the composite system and derive the corresponding field equations by performing the necessary variations. The interaction is introduced at the level of the continuity equations, and the field equations are derived in the flat FLRW background. In Sec.~\ref{sec:dyn_eqn}, we construct the {autonomous system of equations} for both interacting models. A detailed discussion of the observational datasets and the resulting constraints is given in Secs.~\ref{sec:observation_sample} and \ref{sec:results}. In Sec.~\ref{sec:holicow_discussion}, we provide an in-depth analysis of the models' behavior in light of the H0LiCOW data. Finally, a comprehensive summary of our findings is presented in Sec.~\ref{sec:conclusion}.

	\section{ Mathematical framework for interacting $k$-essence with $w$DM fluid \label{sec:basic_frame}}
	
	The action for the $k$-essence field, minimally coupled to the background fluids (wDM, baryons, and radiation), is given by
	\begin{equation}
		S = \int d^4 x \sqrt{-g} \bigg(\frac{R}{2\kappa^2} -  \mathcal{L}_{\phi}(X,\phi) +  \mathcal{L}_{m}\bigg)   ,
	\end{equation}
	where the $k$-essence Lagrangian $\mathcal{L}_{\phi}$ depends on the scalar field $\phi$ and its kinetic term \(X\), and $\mathcal{L}_{m}$ denotes the Lagrangian for the background fluids. We work in natural units $c=\hbar=1$, with the reduced Planck inverse mass defined as $\kappa=\sqrt{8\pi G}$, where $G$ is Newton’s gravitational constant. The pressure of the $k$-essence field is assumed to take the form
	\begin{equation}
		P_{\phi} = -\mathcal{L}_{\phi} = V(\phi) F(X),
	\end{equation}
	where, \(V(\phi)\) is the potential of the field, and \(F(X)\) is a function of the kinetic term, \(X = -\frac{1}{2} g^{\mu\nu} \partial_{\mu}\phi \partial_{\nu} \phi\). 
	
	A potential widely studied in this context, capable of driving late-time cosmic acceleration, is the inverse-square form \cite{Copeland:2006wr,Armendariz-Picon:1999hyi,Jorge:2007zz},
	\begin{equation}
		V(\phi) = \frac{\delta^2}{\kappa^2 \phi^{2}}\ , 
	\end{equation}
	where $\delta$ is the dimensionless constant parameter. For the kinetic function, we adopt the form extensively discussed in the literature,
	\begin{equation}
		F(X) = - X + X^2.
	\end{equation}
	The variation of the action with respect to \(g^{\mu\nu}\) yields the stress-energy tensor of the field,
	\begin{eqnarray}
		T_{\mu \nu}^{(\phi)} = - {\mathcal L}_{,X}\,(\partial_\mu \phi)(\partial_\nu \phi)-g_{\mu \nu}\,{\mathcal L}\, ,
		\label{tphi}  
	\end{eqnarray}
	where ${\mathcal L}_{,X} \equiv \partial \mathcal{L}/\partial X$.
	By comparing with the stress-energy tensor of a perfect fluid, the energy density and pressure of the $k$-essence field are identified as
	\begin{equation}\label{}
		\rho_{\phi} = \mathcal{L} - 2X\mathcal{L}_{,X} \quad \text{and} \quad P_{\phi} = - \mathcal{L}\,.
	\end{equation}
	For the chosen form of \(F(X)\), the density and pressure become
	\begin{equation}
		\rho_{\phi} = -XV+ 3 X^2 V, \ P_{\phi} = V(-X+X^2) \ .
	\end{equation}
	The equation of state (EoS) of the field is then given by
	\begin{equation}
		w_\phi \equiv \frac{P_{\phi}}{\rho_{\phi}} = \frac{(-X+X^2)}{-X+ 3 X^2}\ .
	\end{equation}
	The effective equation of state of the total system can be defined as
	\begin{equation}
		w_{\rm eff} = \frac{P_{\rm tot} = \sum_{i} P_{i}}{\rho_{\rm tot} = \sum_{i} \rho_{i}}\ . 
	\end{equation}
	The continuity equation for the individual components of the Universe can be written as
	\begin{equation}
		\nb_{\mu}^{i} T_{i}^{\mu \nu} = 0\ , 
	\end{equation}
	where the index \({i}\) labels each component.  
	In the present framework, we allow for an interaction between the DM and the $k$-essence field. This is implemented by introducing a source term in their continuity equations. In the spatially flat Friedmann-Lemaître-Robertson-Walker (FLRW) metric, \(ds^2 = -dt^2 + a(t)^2 d\vec{x}^2\), the continuity equations take the form
	\begin{equation}
		\begin{split}
			\dot{\rho}_{m} + 3 H (\rho_{m} + P_{m}) &= Q \ ,\\
			\dot{\rho}_{\phi} + 3 H (\rho_{\phi} + P_{\phi} ) &= -Q \ ,\\
			\dot{\rho}_{r,b} + 3H (1+w_{r,b})\rho_{r,b} &= 0 \ ,
		\end{split}
		\label{continuity_eqn}
	\end{equation}
	{where $\rho_{m}$ and $P_{m}$ denote the energy density and pressure of the matter fluid, respectively; $\rho_{r,b}$ represents the energy densities of radiation and baryons; and $w_{r,b} \equiv P_{r,b}/\rho_{r,b}$ is the corresponding equation of state, and $a$ represents the scale factor. 
		
		In the above equation, the dark matter and $k$-essence field interact through $Q$. For \(Q>0\), energy is transferred from dark energy to the dark matter component. Although the individual equations of motion are modified, the total energy–momentum tensor remains conserved, in agreement with the Bianchi identity. The remaining components are assumed to be minimally coupled to gravity, as early–time observations impose stringent constraints on their energy distribution and behavior \cite{Lee:2006za}.
		
		To study the interaction in more detail, one must specify a form for \(Q\), which may be constant or time-dependent. In the absence of a well–defined microscopic theory of gravity, and given the current lack of observational precision needed to pinpoint the exact nature of these components, it is extremely difficult to determine a unique form of interaction. Therefore, at this stage, it is natural to consider general forms of the interaction that, for nonzero coupling, allow the system to reproduce all known cosmological epochs (radiation $\to$ dark matter $\to$ dark energy), and then examine their effects on cosmic evolution. 
		
		Motivated by this, we consider two general classes of interactions:  
		\begin{eqnarray}
			\text{\bf Model A:}\  Q = 3 H (\alpha \rho_{m} + \xi_1 P_{m} + \beta \rho_{\phi} + \xi_2 P_{\phi}) X^{\gamma}\ , && \label{q1} \\ 
			\text{\bf Model B:}\	Q = 3 H_0 (\alpha \rho_{m} + \xi_1 P_{m} + \beta \rho_{\phi} + \xi_2 P_{\phi}) X^{\gamma}  . && \label{q2}
		\end{eqnarray}
		Here the interaction parameters $(\alpha, \beta, \xi_{1}, \xi_{2}, \gamma)$ are dimensionless constants. In Model A, the interaction is proportional to the time–dependent Hubble parameter, a conventional form that has been widely studied in the literature for various scalar–field and fluid models of dark energy \cite{Wang:2016lxa,Kritpetch:2024rgi}. In Model B, the interaction depends only on the dark matter and dark energy components; dimensional consistency is ensured by including the Hubble constant as the interaction coefficient. We adopt a linear structure for the interaction, combining different dark–sector components that may influence cosmic evolution in distinct ways. In practice, analyzing the full interaction with all terms active simultaneously may obscure the role of individual contributions. Therefore, to better understand the characteristics of each interaction across all redshifts, we activate one coefficient at a time and analyze the resulting dynamics. This approach both simplifies the mathematical structure and enhances numerical stability. Consequently, we categorize both models into submodels, as summarized in Tab.~\ref{tab:int_models}.
		
		A crucial advantage of the $k$-essence field lies in its nonlinear dynamics, which naturally allow the system to explore both phantom and non-phantom regimes without fine-tuning. This makes it a powerful framework for probing the nature of dark energy. For instance, in ref.~\cite{Hussain:2024qrd}, where the $k$-essence field was minimally coupled to cold dark matter, the system exhibited both phantom and non-phantom behavior for certain data combinations, with both states found to be asymptotically stable.  
		\begin{table}
			\begin{tabular}{l | r}
				\hline
				\hline
				{\bf Models} & {\bf Non-zero Parameters} \\
				\hline
				\hline
				{\bf A-I, B-I} & $\alpha \ne 0, \ \gamma \ne 0$\\
				{\bf A-II, B-II} & $\beta \ne 0, \ \gamma \ne 0$\\
				{\bf A-III, B-III} & $\xi_1 \ne 0, \ \gamma \ne 0$\\
				{\bf A-IV, B-IV} & $\xi_2 \ne 0, \ \gamma \ne 0$\\
				
				\hline
				\hline
			\end{tabular}
			\caption{Classes of interacting models.}
			\label{tab:int_models}
		\end{table}
		
		Expanding the continuity equation for the $k$-essence field yields,
		\begin{equation}
			\label{motion1}
			(F_{,X} + 2X F_{,XX})\dot{X}V + 6H F_{,X}X V + (2XF_{,X} - F){V_{,\phi} \dot{\phi}} = -Q,
		\end{equation}
		where the subscript associated with the comma denotes the partial derivative with respect to that quantity. For the inverse-square potential and selected form of \(F(X)\), the evolution equation for the field $\ddot{\phi}$ takes the form:
		\begin{equation}
			\ddot{\phi} = \frac{-Q - 3 H V (-\dot{\phi}^2 + \dot{\phi}^4) + \left(-\dot{\phi}^2/2 + 3 \dot{\phi}^4/4\right) (2 \kappa V^{3/2} \dot{\phi}/\delta)}{(-1 + 3 \dot{\phi}^2) V \dot{\phi}} \ . \label{field_eqn}
		\end{equation}
		The corresponding Friedmann equations are given by}
	\begin{eqnarray}
		3H^2 &=& \kappa^2(\rho_{m} + \rho_{\phi}+ \rho_r + \rho_b),\label{kes_frd1}\\
		2\dot{H} + 3H^2 &=& -\kappa^2 (P_{m} + P_{\phi} + P_r).
	\end{eqnarray}
	In the next section, we reformulate these equations into an autonomous system within the dynamical system stability framework.
	
	\section{Dynamics of the interacting system \label{sec:dyn_eqn}}
	
	To analyze the dynamics of the system, we introduce the following dimensionless dynamical variables:
	\begin{multline}
		x^2 = \frac{\kappa^2 V \dot{\phi}^2}{6 H^2}, \ y^2 = \frac{\kappa^2 V \dot{\phi}^4}{4 H^2}, \ \Omega_{m,r,b, \phi} = \frac{\kappa^2 \rho_{m,r,b, \phi}}{3H^2}, \\ \lambda = \frac{V_{,\phi}}{\kappa V^{3/2}}=-2/\delta\ . 
	\end{multline}
	Here, $\Omega_i$ denotes the fractional energy density of the $i^{\rm th}$ component, while $\lambda$ characterizes the slope of the potential, which remains constant for the inverse square potential. In this work, we treat $\lambda$ as a model parameter instead of $\delta$. The definitions of the dynamical variables are consistent with those in \cite{Hussain:2024qrd}. With these variables, the Hubble constraint leads to the evolution of the dark matter sector as
	\begin{equation}
		\Omega_{m} = 1 - \Omega_{\phi} - \Omega_{r} - \Omega_{b}  \ ,
		\label{const_dyn}
	\end{equation}
	where $\Omega_{\phi} = -x^2 + y^2$. The physically viable solutions are those for which the fractional energy densities satisfy $0 \leq \Omega_{i} \leq 1$. 
	The effective equation of state (EoS) for the interacting system is given by 
	\begin{equation}
		w_{\rm eff} = \frac{P_{\rm tot}}{\rho_{\rm tot}} = \frac{-2 \dot{H}}{3H^2} -1\ , 
	\end{equation}
	where the Hubble derivative expands to
	\begin{equation}
		\frac{\dot{H}}{H^2} = -\frac{3}{2} \left(w_m \Omega_{m} + \frac{1}{3}\Omega_{r} - x^2 + \frac{y^2}{3} + 1\right)\ .
	\end{equation}
	{
			Besides the equation-of-state (EoS) parameter, an adiabatic sound speed is usually defined at the perturbation level, which characterizes the propagation speed of scalar perturbations. The positivity of this quantity is a necessary condition to avoid ghost and gradient instabilities in the perturbed system. In the present case, the sound speed takes the form\footnote{
				The definition of the sound speed for the interacting system considered here coincides with that of the minimally coupled case. Since our analysis is restricted to the background level, the positivity of the sound speed for non-canonical fields provides an essential consistency check related to the stability of the background evolution. Accordingly, the sound speed is required to satisfy $0 \le c_s^2 \le 1$ throughout the cosmic evolution.}
			\cite{Garriga:1999vw,Linton:2017ged,Valiviita:2008iv,Piattella:2013wpa}	
	\begin{equation}
		\cs  = \frac{\partial P_{\rm tot}/\pp X}{\pp \rho_{\rm tot}/ \pp X} = \frac{-1+ 2/3 \  y^2/x^2}{-1+ 2 y^2 /x^2} \ .
		\label{sound_speed}
		\end{equation}}
	In this study, we consider the dark matter equation of state to satisfy $w_m \neq 0$, while the EoS values for radiation and baryons are taken to be $1/3$ and $0$, respectively. We now proceed to construct the autonomous system of equations for the composite system as
	\begin{eqnarray}
		\frac{\dot{x}}{H} & =& \frac{1}{2} \lambda \sqrt{6} x^2 + x \frac{\ddot{\phi}}{\dot{\phi} H} - x\frac{\dot{H}}{H^2} \label{x_prime} \ ,\\
		\frac{\dot{y}}{H} & =& \frac{\sqrt{3} \lambda y x}{\sqrt{2} } + \frac{2 y \ddot{\phi}}{\dot{\phi} H} - y \frac{\dot{H}}{H^2} \label{y_prime}\ ,\\
		\frac{\dot{\Omega}_{r}}{H} & =& - 4 \Omega_{r} - 2 \Omega_{r} \frac{\dot{H}}{H^2} \ , \label{omega_r_prime}\\
		\frac{\dot{\Omega}_b}{H} & =& -3 \Omega_{b} - 2 \Omega_{b} \frac{\dot{H}}{H^2} \ , \label{b_prime}
	\end{eqnarray}
	where $\dot{()} \equiv d()/dt$, and the new time variable is defined as $dN = H dt$. Using Eq.~\eqref{field_eqn}, the expression for $\ddot{\phi}/(\dot{\phi} H)$ becomes
	\begin{eqnarray}
		\frac{\ddot{\phi}}{\dot{\phi} H} &&= \frac{-\frac{Q}{VH} -  (-2y^2/x^2 + 4/3 y^4/x^4) + \left(-\frac{ y^2}{3x^2} + \frac{y^4}{3 x^4}\right) (-\lambda \sqrt{6} x)}{(-1+2y^2/x^2)\left(\frac{2 y^2}{3x^2}\right)}\ .\nonumber \\ 
	\end{eqnarray}
	Here, $Q/(VH)$ is a dimensionless quantity that can be expressed in terms of the predefined variables for the selected models. For Model A, this takes the form
	\begin{multline}
		{\textbf{\text{Model A:}}}\	\frac{Q}{VH} = \bigg(\alpha \Omega_{m} \frac{y^2}{x^4} + \xi_{1} w_m \Omega_{m} \frac{y^2}{x^4}\\ + \beta \Omega_{\phi} y^2/x^4 + 3 \xi_{2} F\bigg) \left(\frac{y^2}{3x^2}\right)^{\gamma} \ .
	\end{multline}
	Because of the multiplicative factor $H$, the interaction is expressed entirely in terms of the predefined variables without introducing additional ones. Thus, the phase space remains four-dimensional, with the Hubble constraint equation Eq.~\eqref{const_dyn} providing a stringent condition on the system. We do not discuss the stability of the system through critical points at this stage; instead, in the next section, we constrain the model parameters with cosmological data and analyze the qualitative behavior of the model.  
	
	In Model B, an additional factor of $H^{-1}$ appears in the $Q/(VH)$ term. To properly close the autonomous system, we define an additional dimensionless variable
	\begin{equation}
		h \equiv H/H_0\ ,
	\end{equation}
	which modifies the interaction function to
	\begin{multline}
		{\textbf{\text{Model B:}}}\	\frac{Q}{VH} = \frac{1}{h}\bigg(\alpha \Omega_{m} \frac{y^2}{x^4} + \xi_{1} w_m \Omega_{m} \frac{y^2}{x^4}\\ + \beta \Omega_{\phi} y^2/x^4 + 3 \xi_{2} F\bigg) \left(\frac{y^2}{3x^2}\right)^{\gamma} \ ,
	\end{multline}
	This extends the dimensionality of the phase space from $4 \to 5$. Since the new variable is time dependent, the time derivative of $h$ must be included in the autonomous system, only for Model B:
	\begin{equation}
		\frac{	\dot{h}}{H} = h \frac{\dot{H}}{H^2} \ .
		\label{h_prime}
	\end{equation}
	Thus, the dynamics of the system are described by Eqs.~(\ref{x_prime})--\eqref{b_prime} together with Eq.~\eqref{h_prime}. It is customary to investigate stability by analyzing the critical points of the autonomous system\footnote{Critical points are obtained by equating the right-hand side of the autonomous equations to zero, i.e., $\vec{x}_i = 0$, where $i$ runs from $1$ to the number of independent equations \cite{Bahamonde:2017ize,Hussain:2024qrd}.}. For the present model, one such coordinates of the critical point corresponds to \(h = 0\), for which the \((x' \equiv \dot{x}/H)\) and \(y'\) from Eqs.~\eqref{x_prime} and \eqref{y_prime} diverges whenever \((x , y \ne 0)\). This critical point $h \to 0$ corresponds to the epoch where $H \ll H_0$, i.e., the far future of the universe. In contrast, during the past epoch when $H \gg H_0$, we have $h \neq 0$, and the system remains free of divergences. Therefore, the model does not exhibit pathological solutions as long as we restrict our analysis to the past evolution of the universe, which is the regime probed by current cosmological observations.  
	
	The irregularity in the future behavior can be addressed by redefining the time variable $dN = H dt$ as
	\begin{equation}
		dN \mapsto h d \bar{N} \ ,
	\end{equation}
	which cancels the additional $h^{-1}$ in the autonomous equations. This method is well established within the dynamical systems framework for studying stability \cite{Bahamonde:2017ize}. We refer interested readers to earlier works that explore this approach in detail \cite{Bouhmadi-Lopez:2016dzw,Alho:2020cdg,Das:2019ixt}.  
	
	As our main goal is to constrain the model parameters from an observational perspective, we shall present the solutions using the original time variable $N = \ln a$, which relates to the redshift \((z)\) as
	\begin{equation}
		N = -\ln (1+z)\ . 
	\end{equation}

	\section{Observational Data Sets \label{sec:observation_sample}}
	In this section, we briefly describe the data sets employed to constrain the model parameters.

	\begin{itemize}
		\item \textbf{CC Data:} This data set consists of $32$ model-independent Hubble parameter measurements spanning the redshift range $z \in [0.07, 1.965]$. We use the covariance matrix constructed in \cite{Moresco:2012jh,Moresco:2015cya,Moresco:2016mzx} to evaluate the likelihood\footnote{The Python code to construct the covariance matrix for the $15$ highly correlated samples is available at \url{https://gitlab.com/mmoresco/CCcovariance}}. 
		
		\item \textbf{PP Data:} This sample corresponds to Type Ia supernovae, containing $1701$ observations from the Pantheon+ compilation, covering $z \in [0.001, 2.26]$ \cite{Brout:2022vxf}. We use the non-SH0ES calibrated subset by applying the filter $z>0.01$, reducing the number of data points to $1590$. The observable is the apparent magnitude,  \(\rm m_{\rm obs} = 5 \log (D_L/Mpc) + 25 + M_{\rm B}\), where $M_B$ is the nuisance parameter (absolute magnitude) and $D_L$ is the luminosity distance. We analytically marginalize $M_B$ following the prescription in \cite{Goliath:2001af}, also implemented in the \texttt{Cobaya} repository\footnote{Likelihood estimation code: \url{https://github.com/CobayaSampler/cobaya/blob/master/cobaya/likelihoods/sn/pantheonplus.py}.}. We refer to this data set as ``PP''. 
		
		\item \textbf{DES Data:} This sample includes Type Ia supernovae from the Dark Energy Survey (\textbf{DES-SN5YR}), consisting of $1829$ data points \cite{DESI:2024mwx}, hereafter denoted as ``DES''. The observable is the distance modulus $\mu$, with the nuisance parameter $M_B$ pre-calibrated. The likelihood is computed by marginalizing $M_B$ using the code available in the \href{https://github.com/des-science/DES-SN5YR/blob/main/5_COSMOLOGY/Dovekie_cosmosis_likelihood.py}{DES-SN5YR module}. Throughout the analysis, PP and DES samples are not combined, as both catalogs share some overlapping data points. 
		
		\item \textbf{DESI BAO:} This data set corresponds to Baryon Acoustic Oscillation (BAO) measurements from the Dark Energy Spectroscopic Instrument (DESI) Release II \cite{DESI:2025zgx,DESI:2019jxc}, an improved version of DESI DRI. The observables are the ratios $\{D_M/r_d, D_H/r_d, D_V/r_d\}$, where $D_M$ is the comoving angular distance, $D_H$ is the comoving Hubble distance, $D_V$ is the spherically averaged distance, and $r_d$ is the sound horizon at the drag epoch \cite{eBOSS:2020yzd}. The sound horizon is computed as 
		\begin{equation}
			r_{d} =\int_{ z_{d}}^{\infty} \dfrac{3 \times 10^{5} d\ti{z}}{H(\ti{z}) \sqrt{3\left(1 + \frac{3 \Omega_{b}h^2}{4 \Omega_{\gamma}h^2(1+\ti{z})}\right)}} \ ,
			\label{sound_distance}
		\end{equation}
		where $z_d$ is estimated using the Hu--Sugiyama fitting formula \cite{Hu:1995en},
		\begin{equation}
			\begin{aligned}
				& z_{\mathrm{d}}=1345 \frac{\left(\Omega_{\mathrm{m}} h^2\right)^{0.251}\left[1+b_1\left(\Omega_{\mathrm{b}} h^2\right)^{b_2}\right]}{1+0.659\left(\Omega_{\mathrm{m}} h^2\right)^{0.828}}\ , \\
				& b_1=0.313\left(\Omega_{\mathrm{m}} h^2\right)^{-0.419}\left[1+0.607\left(\Omega_{\mathrm{m}} h^2\right)^{0.674}\right] \ , \\
				& b_2=0.238\left(\Omega_{\mathrm{m}} h^2\right)^{0.223} \ .
			\end{aligned}
		\end{equation}
		Here, $\Omega_m$ denotes the total matter density (dark matter + baryons), $\Omega_b$ is the baryon density, and $h \equiv H_0/100$\footnote{Note that this $h$ differs from the dynamical variable $h$ introduced in the autonomous system. In the present context, $h$ appears only in combination with the fractional energy densities and represents the normalized Hubble constant.}. The photon density is fixed to $\Omega_\gamma h^2 = 2.47 \times 10^{-5}$.  
		
		\item \textbf{PLA Data:} This data set is a compressed version of the full Planck likelihood \cite{Planck:2018vyg}, reported in \cite{Arendse:2019hev}. We adopt the extended four-parameter likelihood $\{100 \Omega_b h^2, 100 \theta_{*}, R, \Omega_{\rm dm}h^2\}$ with its correlation matrix. The extended prior space efficiently constrains models beyond $\Lambda$CDM. The parameters include the shift parameter $R = \sqrt{\Omega_m} H_0 D_A(z_*)/c$ and the angular scale $\theta_{*} = r_s(z_*)/D_A(z_*)$, where $z_*$ is the recombination redshift, $r_s(z_*)$ the sound horizon at recombination, and $D_A$ the comoving angular diameter distance,
		\begin{equation}
			D_A = c \int_{0}^{z}\frac{dz}{H(z)}\ . 
		\end{equation}
		
		\item \textbf{BBN Data:}  We also include Big Bang Nucleosynthesis (BBN) constraints using the \texttt{montepython} implementation\footnote{\url{https://github.com/brinckmann/montepython_public/tree/3.6/data/bbn}} \cite{Cooke:2018qzw}. The likelihood is computed following the code provided in \href{https://github.com/brinckmann/montepython_public/tree/3.6/montepython/likelihoods/bbn_omegab}{this module}. We use the \textit{Primat} measurement of primordial abundances \cite{Pitrou:2018cgg}.
		
	\end{itemize}
	We evaluate the joint likelihood using the following data combinations:
	\begin{equation*}
		\rm  (I). \ CC+BAO+PLA+PP \ (II). \  CC+BAO+PLA+DES \ .
	\end{equation*}
	Here, the combination CC+BAO+PLA serves as the baseline, providing essential constraints on extensions beyond $\Lambda$CDM. Both combinations capture the key features of the models and yield constraints relevant to different cosmological epochs. The total log-likelihood is defined as 
	\begin{equation}
		\ln	(\mathcal{L}_{\rm tot}) = -\frac{1}{2} \chi^{2}_{\rm tot} \ ,
	\end{equation}
	with
	\begin{equation}
		\chi_{\rm tot}^2 = \chi^{2}_{\rm CC} + \chi^{2}_{\rm BAO} + \chi^{2}_{\rm PLA} +\chi^{2}_{\rm PP} (\chi^{2}_{\rm DES} )\ . 
	\end{equation}
	The BBN likelihood will be discussed in the next section. Posterior distributions are obtained via Bayes’ theorem, requiring a careful choice of prior ranges. Based on initial numerical evolution, we identify extended parameter ranges that yield consistent solutions across cosmic epochs, and adopt uniform priors as listed in Tab.~\ref{tab:prior_range}. For sampling and likelihood evaluation, we employ the nested sampler \texttt{PolyChord}, which is well suited for high-dimensional parameter spaces compared to affine-invariant MCMC samplers such as \texttt{emcee} \cite{Handley:2015fda,Handley:2015vkr,Foreman-Mackey:2012any}. The chains are analyzed using \texttt{GetDist} to extract the best-fit values \cite{Torrado:2020dgo,Lewis:2019xzd}. Finally, we compare our models against flat $\Lambda$CDM using information criteria such as the Akaike Information Criterion (AIC) and Bayesian Information Criterion (BIC) \cite{Akaike:1974vps,wenren2016marginal,Trotta:2008qt}, following the standard methodology described in \cite{Hussain:2024jdt}. 
	
	\section{Results \label{sec:results}}
	In this section, we present the results obtained by implementing the models described in Tab. \ref{tab:int_models} with the considered combination of the data sets. We choose a uniform prior range on most of the variables except radiation density, where we adopted a gaussain prior. The prior range for all the models are listed in Tab. \ref{tab:prior_range}. We intentionally select a positive range for dark matter equation of state, as a negative value yields a negative sound speed at the perturbation level, which in many previous studies artificially set to zero \cite{Li:2025eqh}. Hence, to avoid the  instability (which does not arise in the background level), we select a physically viable range \(w_m \in [0, 0.01]\). 
	
	\begin{table}
		\centering
		\begin{tabular}{l r }
			\hline
			Parameters & Range \\
			\hline
			\multicolumn{2}{c}{\bf \boldmath Model A: \(\gamma =1\), \& B: $\gamma = -1$ } \\
			\hline
			{\boldmath $\Omega_{\phi}$} & [0.5, 0.9]  \\
			
			{\boldmath $H_0$ } & [30,100]  \\
			
			{\boldmath $x_0$ } & $[1.03,\ 1.37]$ \\
			
			{\boldmath $\lambda$ } & $[10^{-5},\  0.3]$ \\
			
			{\boldmath $w_m$} &  [0,0.01] \\
			
			{\boldmath $\Omega_b h^2$} &  [0,0.05] \\
			
			{\boldmath $\Omega_{r}$} &  $\mathcal{N}$[$9.1 \times 10^{-5}$, $10^{-6}$] \\
			
			{\boldmath $\alpha$} & [$10^{-12}$, $1$]  \\
			
			{\boldmath $\beta$} & [$10^{-14},\ 1$] \\
			
			{\boldmath $\xi_{1}$} & [$10^{-12}$, $1$]\\
			
			{\boldmath $\xi_{2}$} & [$-10^{-12}$, $-1$] \\			
			
			\hline
			\hline
		\end{tabular}
		\caption{The prior range of the model parameters.}
		\label{tab:prior_range}
	\end{table}

	\subsection{Model A, $\gamma =1$}

	We categorize Model A into four sub-models, labeled I–IV. This framework incorporates both dark matter and dark energy components, with the assumption that only one interaction channel is active at a time\footnote{ We have also explored scenarios in which more than one interaction parameter is simultaneously non-zero. In such cases, we find that the current observational data are unable to tightly constrain the model parameters due to strong degeneracies among the interaction terms, which prevents drawing robust conclusions about the physical behavior of the interacting system at the background level. We therefore restrict our analysis to minimal, representative interaction scenarios.}. The interaction term includes an additional kinetic contribution of the form $X^{\gamma}$, where we fix $\gamma = 1$ to retain effects at linear order. Our analysis shows that deviations from $\gamma = 1$ lead to {numerical} instabilities, preventing the model from reproducing the standard cosmological sequence (Radiation $\to$ Matter $\to$ Dark Energy). \footnote{ The parameter $\gamma$ can in principle be treated as free and explored within a range (e.g., $-2 < \gamma < 2$). However, for values of $\gamma$ other than $\gamma = 1$ (Model~A) and $\gamma = -1$ (Model~B), the autonomous system becomes increasingly stiff at high redshift, leading to unstable background solutions. Although stable evolution may occur for finely tuned initial conditions, such solutions are not observationally robust and introduce significant bias in the MCMC analysis due to strong nonlinearities. We therefore restrict our study to these representative cases.}
		
		The marginalized posterior distributions of the parameters, after integrating over the interaction couplings $\{\theta_i\} \equiv \{\alpha, \beta, \xi_1, \xi_2\}$, are shown in Fig. \ref{fig:corner_modelab_sn}. The corresponding best-fit values (68\% confidence level) are summarized in Tab. \ref{tab:model_A+B_best_fit_sn}.
	
	\begin{itemize}
		\item \textbf{Model A I:} This case corresponds to $Q \propto H \rho_{\rm m} X$.  The best-fit Hubble constant is $H_0 \approx 67.1$ km/s/Mpc (BASE+PP) and $H_0 \approx 69.6$ km/s/Mpc (BASE+DES), in close agreement with the flat $\Lambda$CDM values (Tab. \ref{tab:lcdm_holicow}), and in strong tension with the SH0ES result $H_0 = 73.6 \pm 1.1$ km/s/Mpc \cite{Brout:2022vxf}. The interaction parameter $\alpha$ is strongly suppressed, $\alpha \sim \mathcal{O}(10^{-10})$, while the potential slope is $\lambda \sim 0.007$. The model yields field energy density of about $\sim 69.2 \%$. The dark matter EoS is small but positive, $w_m \sim 10^{-4}$, consistent with the near-pressureless nature found in earlier works \cite{Kumar:2012gr,Pan:2022qrr}.			
		
		To understand the impact of the interaction on the cosmological epoch, we evolve the cosmological parameters such as the density parameters, equation of state parameters, and the interaction quantity \((\Omega_{(\phi, m, b, r)}, w_{\phi}, w_{\rm eff}, Q/(VH))\) against $N = - \ln (1+z)$, where \(z\) represents the redshift. The illustration is shown in Fig.~\ref{fig:evo_modelA_sn}. The figure reveals that, due to the diminishing magnitude of $\alpha$, at low redshift the overall interaction quantity becomes significantly smaller. However, it increases exponentially with increasing redshift. This is because of the dependence on the Hubble parameter, the dark matter energy density, and kinetic function $X$, which increase with redshift. Therefore, the interaction model does not exhibit a significant impact in the lower-redshift regime compared to the higher-redshift regime, which also demonstrates that the model only slightly pushes the value of the Hubble parameter.  
		
		Due to the very small magnitude of $\alpha$, the density parameters show consistent behavior across the entire redshift range, clearly exhibiting different epochs at distinct scales. As pointed out in earlier investigations, for this class of interacting models, if the interaction parameter does not take a small value, the model develops instabilities at the perturbation level \cite{Wang:2016lxa}.  
		
		Unlike the potential-driven quintessence field, whose energy density takes the form \(\rho_{\phi} = \frac{\epsilon}{2} \dot{\phi}^2 + V(\phi)\), where as long as $\epsilon > 0$, the equation of state converges to $-1$ in the late-time epoch and does not cross the phantom divide unless the switching parameter is set to $\epsilon = -1$, the current model features both phantom and non-phantom solutions depending on the initial condition \(x_0\) and the potential parameter $\lambda$. In the evolution plot, we see that the individual EoS corresponding to the scalar field, $w_\phi$, tracks the background fluid EoS at high redshift where it takes the value \(w_\phi = 1/3\). As the system transitions to the matter-dominated epoch, it starts evolving towards negative values. However, the field energy density remains subdominant up to $N=-1$, while its EoS continues decreasing toward negative values.  
		
		Nevertheless, we see that the effective equation of state becomes $1/3$ during the radiation-dominated epoch, and approaches zero during the matter-dominated phase, where $\Omega_{\rm m}$ dominates over the rest of the components. Near the redshift $N = -2$, while \(w_{\rm eff}\) remains close to zero, the field EoS \(w_\phi\) approaches very close to $-1$, and asymptotically saturates there, representing a non-phantom regime. When extrapolated to the late-time future epoch, both the EoSs \(w_{\rm eff}\) and \(w_\phi\) coincide and approach $-1$, representing a de Sitter universe. On extrapolation to the future epoch $N>0$, the field density $\Omega_{\phi}$ saturates to $1$ with effective EoS $-1$, thereby rendering a non-phantom solution and reproducing all the observable epochs in the past.   
		
		\item {\bf Model A II:} This corresponds to the interaction form where the coupling is proportional to the dark energy density, i.e., \(Q \propto H \rho_{\phi} X\). In the previous model, where the interaction was proportional to the dark matter energy density, its effect at higher redshifts was significantly stronger. In contrast, the dark energy density in the past is non-zero but extremely small compared to that of dark matter. Therefore, we analyze this model to examine its impact on the evolution of the Hubble parameter across redshift. The model yields $H_0 \simeq 67.0$ km/s/Mpc for the PP data set and $H_0 \simeq 69.64$ km/s/Mpc for the DES data set. The interaction parameter $\beta$ becomes extremely small, taking values in the range $0 < \beta < 10^{-12}$. It should be noted that $\beta$ is always constrained to be positive, since negative values render the differential system unstable. The remaining parameter values are nearly consistent with the previously examined model. For the best-fit values, we plotted the interaction function in Fig.~\ref{fig:evo_modelA_sn}, which shows that the interaction magnitude at low redshift is much smaller compared to the previous model. At higher redshift, however, the interaction remains relatively weaker than in the preceding case. A very small value of $\beta$ suggests that this type of interaction, where the dark energy density dominates only at low redshifts, may destabilize the system, making the realization of distinct cosmological epochs non-feasible. The evolution of the density parameters and the EoS shows a similar trend to that of the previous model.

		\item {\bf Model A III:} In this model, the interaction function is proportional to \(Q \propto H P_{m} X\). Since \(w_m \ne 0\), the interaction becomes non-zero. The model yields $H_0 \simeq 67.0$ km/s/Mpc for the PP data set and exceeds $H_0 \simeq 69.61$ km/s/Mpc for the DES data set. The remaining parameters take values similar to those of the previous model. The evolution of the interaction function shows that the interaction remains significantly smaller, of the order \(\mathcal{O}(10^{-8})\), at low redshifts. Overall, the model exhibits a cosmological behavior similar to that of the previous case. 
		
		\item {\bf Model A IV:} In this model, the interaction function is proportional to \(Q \propto H P_{\phi} X\). Due to its dependence on the field pressure, the interaction can transit from positive to negative values, reflecting the behavior of the scalar field equation of state. From the continuity equation, Eq.~\eqref{continuity_eqn}, we see that as long as \(Q > 0\), dark energy loses energy to dark matter. In the previous models, DE continuously transfers energy to DM throughout cosmic evolution. In contrast, due to the transitional behavior of \(w_\phi\), this model allows energy to flow from DM to DE or from DE to DM at different redshifts.
		
		Evaluating the interaction function for the best-fit values (Fig. \ref{fig:evo_modelA_sn}), we find that the interaction changes from negative to positive near \(N \sim -3.0\). The positive value at low redshift indicates energy flow from DE \(\to\) DM, while for \(N < -3\), the interaction becomes negative, indicating energy flow from DM \(\to\) DE. Due to this transitional behavior, the magnitude of the interaction is three to four orders of magnitude higher than in the other models at low redshift. At higher redshift, the interaction does not increase rapidly and saturates around \(10^{6}\), successfully reproducing all cosmic epochs. The prior range for the interaction parameter \(\xi_2\) is chosen to be negative because positive values produce instabilities and prevent the model from reproducing the observable epochs. The model yields \(H_0\) similar to the other models, while the dark matter EoS is marginally higher, \(w_m \sim 0.0003\), compared to the rest of the cases.
		
	\end{itemize}
	
	In summary, all the considered interacting models yield values of \(H_0\) similar to those of flat $\Lambda$CDM; however, a mild increase in \(H_0\) is observed compared to the non-interacting case. Therefore, the DM–DE interaction can slightly raise the expansion rate. For the chosen data sets, particularly the PP sample, the results remain in significant tension with the SH0ES measurement. On the other hand, models using the DES supernova samples consistently yield higher \(H_0\) values than PP, thus showing tension with the PP data. Among the four models, Model A-IV exhibits a significantly higher interaction magnitude at low redshift, where dark energy transfers energy to dark matter. As the system evolves toward the past epoch, the energy flow reverses, and dark matter begins transferring energy to dark energy. Although this transitional behavior arises from the internal structure of the scalar field, the interaction eventually freezes in the asymptotic past and future, leading to a stable evolution.
	
	{Additionally, since all the models considered exhibit non-phantom behavior during the current and future epochs, we plot the evolution of the sound speed defined in Eq.~\eqref{sound_speed} for both datasets in Fig.~\ref{fig:sound_speed_sn}. This illustrates that the sound speed of the non-canonical field remains positive throughout the entire cosmic evolution, providing a necessary consistency check related to the absence of ghost-like behavior. 	\\	
		We emphasize, however, that the positivity of the sound speed alone does not constitute a complete perturbative stability analysis. As a full treatment of the perturbation equations for the interacting system is beyond the scope of the present work, we refrain from making definitive claims regarding perturbative stability. A detailed perturbation-level analysis of the models considered here will be pursued in future work.
	 }
	 
	 	{The reader should note that the background evolution shown in Fig.~2, as well as in the remaining evolution figures, is presented down to $N \simeq -13$. We have explicitly verified that extending the integration further into the past does not lead to any qualitative change in the radiation-dominated behavior or in the overall background dynamics. In particular, the interaction terms become effectively negligible at high redshifts, which explains the near-identical evolution across different interacting models at early times. This behavior is closely tied to the requirement that the models reproduce the observed sound horizon scale at decoupling to be close to $147$ Mpc, which enforces a standard radiation-dominated expansion history at high redshift when constrained by the compressed Planck data. As a result, the impact of the interaction parameters on the background evolution is naturally suppressed and difficult to distinguish at the level of the background dynamics.

	\subsection{Model B, $\gamma =-1$}

	To constrain the parameters of the current interaction form, we initially fix the parameter \(\gamma = -1\). The choice of \(\gamma\) is motivated by the model's numerical behavior under several initial conditions. For \(\gamma = 1\), the interaction exhibits a trend similar to the previous models, monotonically increasing with redshift. In contrast, for \(\gamma = -1\), the interaction becomes less sensitive at high redshift and shows a more pronounced effect at low redshift. The variations of the interaction for different models are illustrated in Fig. \ref{fig:evo_modelB_sn}. The prior ranges of the model parameters are listed in Tab. \ref{tab:prior_range}. The best-fit values are summarized in Tab. \ref{tab:model_A+B_best_fit_sn}, and the corresponding marginalized posterior distributions are shown in Fig. \ref{fig:corner_modelab_sn}.
	
	Nearly all models yield a dark energy density of \(\sim 69.0\%\) for the PP data set and \(\sim 69.4\%\) for the DES data set. A mild increase in \(H_0\) is observed with the PP data set: \(H_0 \sim 67.18\) km/s/Mpc for Model II and \(H_0 \sim 67.13\) km/s/Mpc for Model III, slightly higher than in the previous models. For the DES data set, nearly all models produce similar \(H_0\) values. The dark matter equation of state remains small, in the range \((1.8-2.0) \cdot 10^{-4}\). As \(\lambda\) becomes very small, the scalar field EoS asymptotically approaches \(-1\) during the late-time epoch. All models successfully reproduce the distinct cosmological phases without exhibiting phantom behavior, even without imposing any bias in the system.
	
	The interaction becomes particularly significant during the late-time epoch. In Fig. \ref{fig:evo_modelB_sn}, Models I and III show a diminishing interaction at very low and very high redshift, while reaching a significantly higher value during the matter-dominated epoch at intermediate redshift. This indicates that the current functional form of the interaction could play an important role in structure formation and may lead to deviations from \(\Lambda\)CDM predictions when tested against large-scale structure observations.
	
	In contrast, the interaction magnitude for Models II and IV increases at low redshift, where the dark energy density is higher. For \(z < 2\), the interaction is of order \(\mathcal{O}(10^{-7})\). The interaction remains positive for Model B-II, but crosses zero for Model B-IV around \(N \sim -4\). Unlike Models A-I to A-III, these models do not show abrupt growth at high redshift but instead attain higher values at intermediate redshift (Models I and III) and low redshift (Models II and IV). {Additionally, we plot the sound speed for each model in Fig.~\ref{fig:sound_speed_sn}, where it can be seen that the sound speed remains positive throughout the entire evolution, providing a necessary consistency check related to the absence of ghost-like behavior.}

	The statistical measures of the models are summarized in Tab. \ref{tab:stat_aic_bic}. The minimum chi-squared values for all models are slightly smaller than those of the flat $\Lambda$CDM model for the PP data set, with Model A-IV providing the best fit among all. For the DES data set, the models yield slightly higher chi-squared values, although the value for Model A-IV is approximately equal to that of $\Lambda$CDM. This indicates that the interacting models provide an excellent fit to the considered data sets.
	
	We also report the AIC and BIC values, which are naturally higher than those of the reference model due to the additional degrees of freedom associated with the scalar field models. Among the interacting models, Model A-IV consistently outperforms all others based on both AIC and BIC criteria.

	\begin{figure*}
		\includegraphics[scale=0.8]{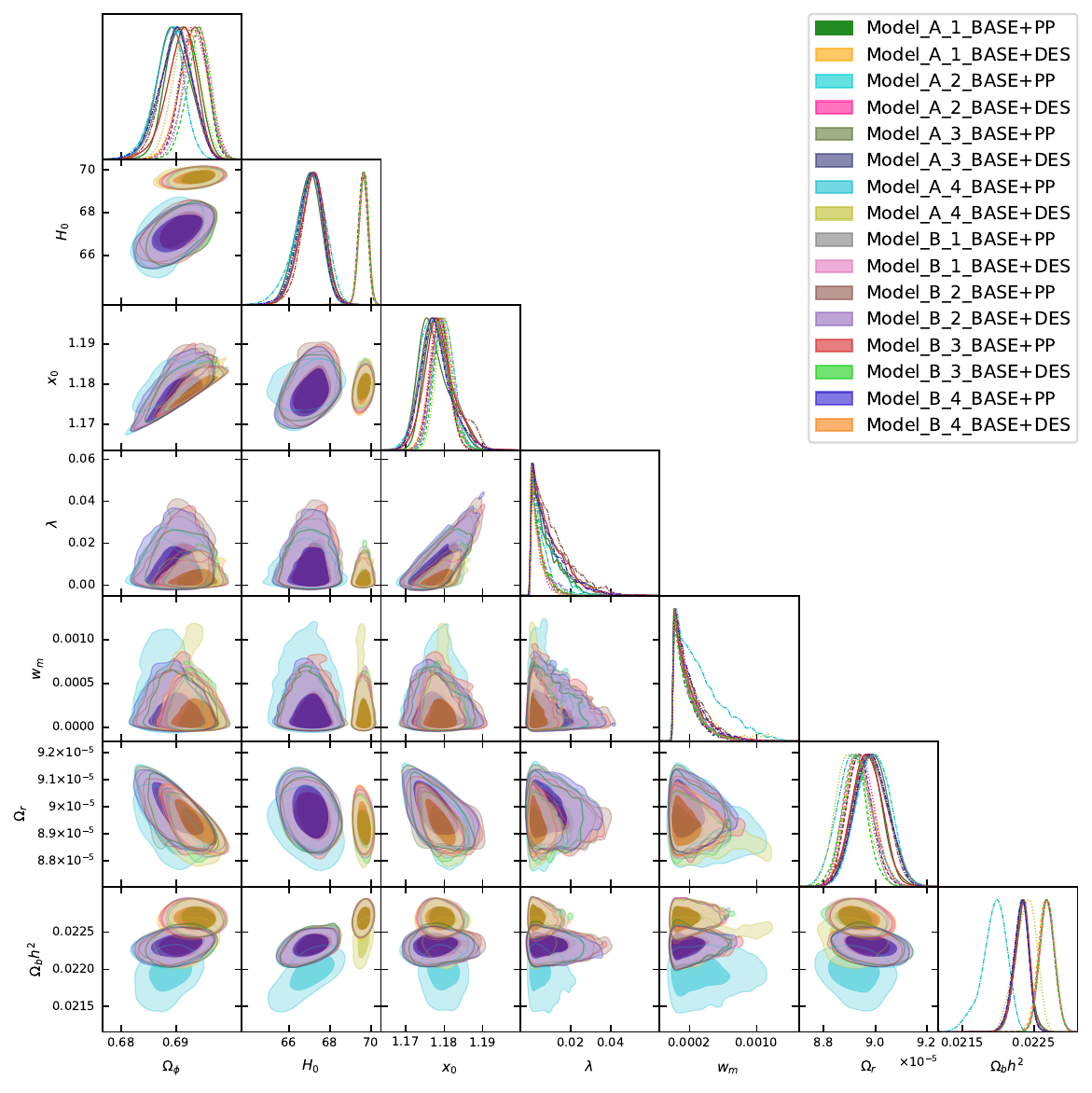}
		\caption{Posterior distributions of the model parameters for Models A and B, with the interaction parameter marginalized, using the supernova datasets.}
		\label{fig:corner_modelab_sn}
	\end{figure*}
	
	\begin{figure*}
		\includegraphics[scale=0.5]{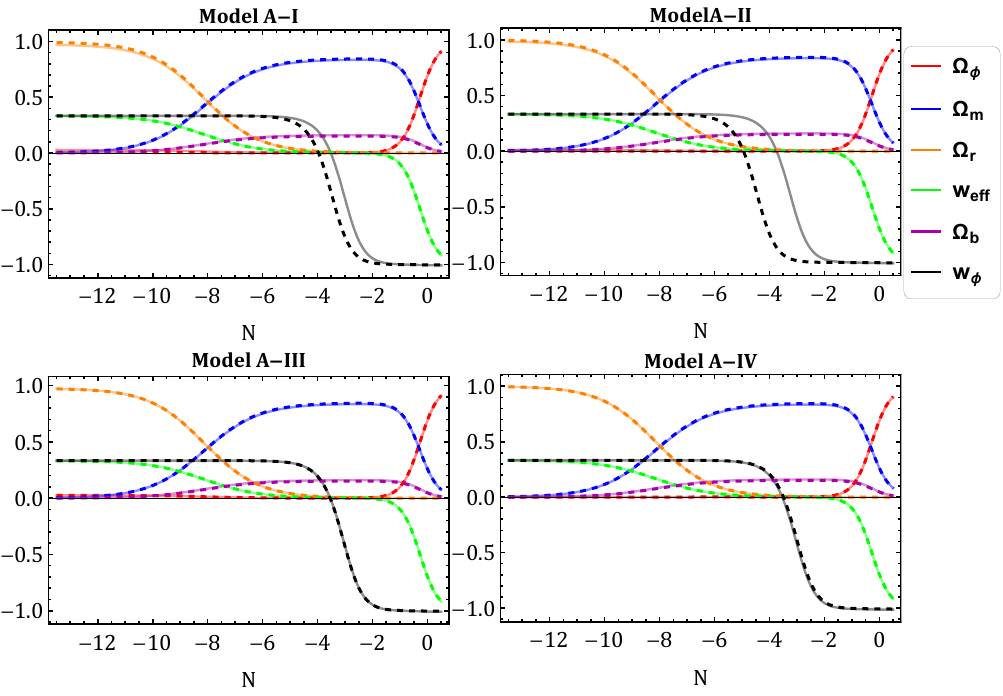}
		\includegraphics[scale=0.5]{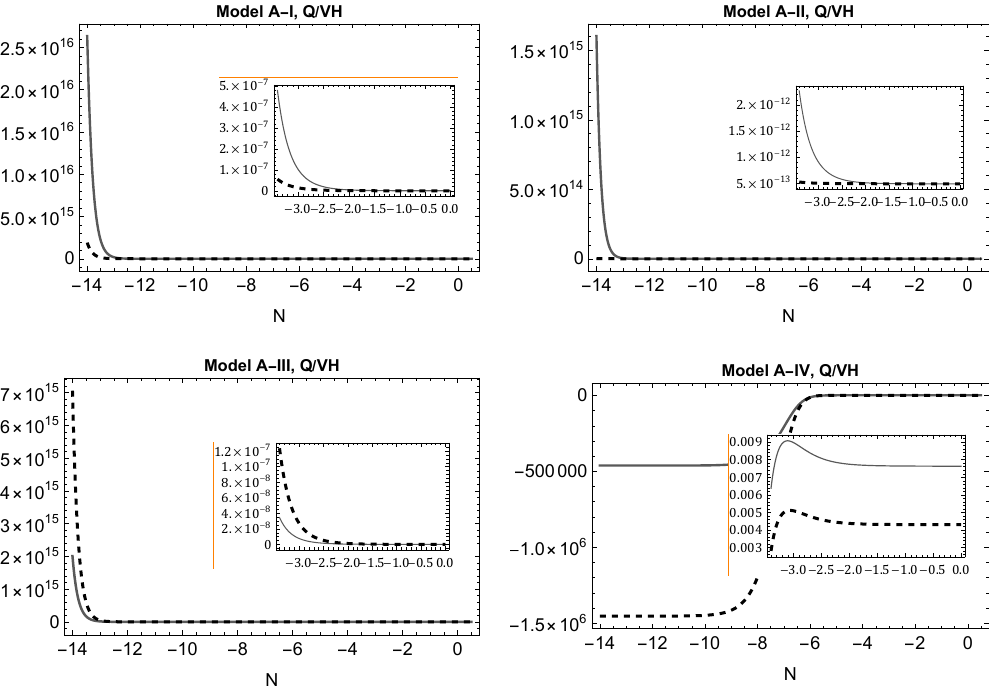}
		\caption{Evolution of the cosmological parameters for Model A, corresponding to the best-fit values obtained from MCMC. Solid and dashed lines represent the best-fit values for BASE+PP and BASE+DES datasets, respectively.}
		
		\label{fig:evo_modelA_sn}
		
	\end{figure*}

	\begin{figure*}
		\includegraphics[scale=0.5]{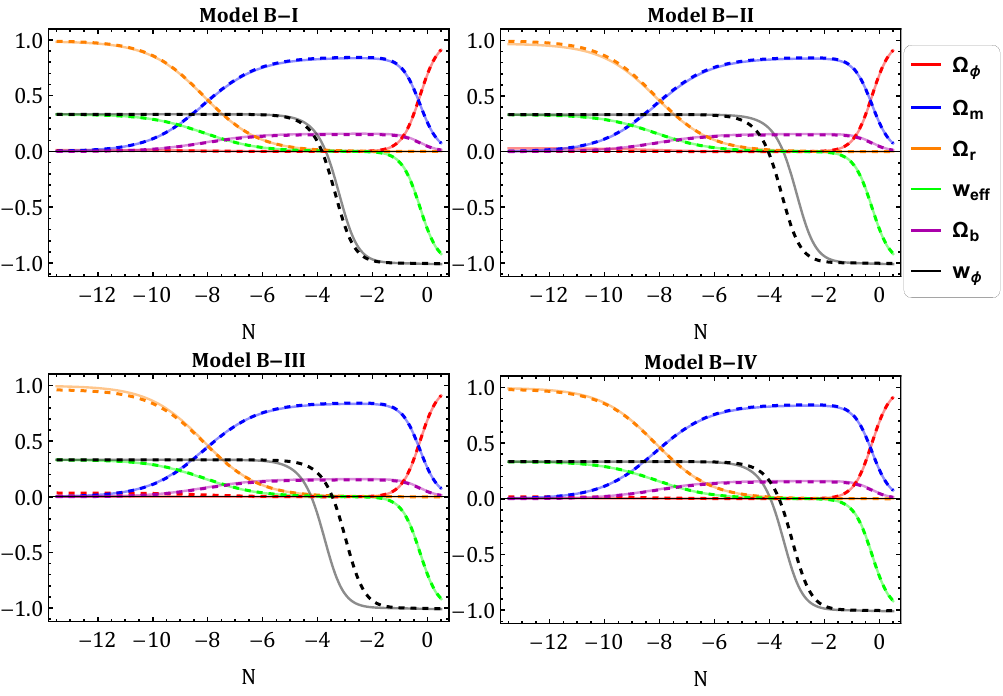}
		\includegraphics[scale=0.5]{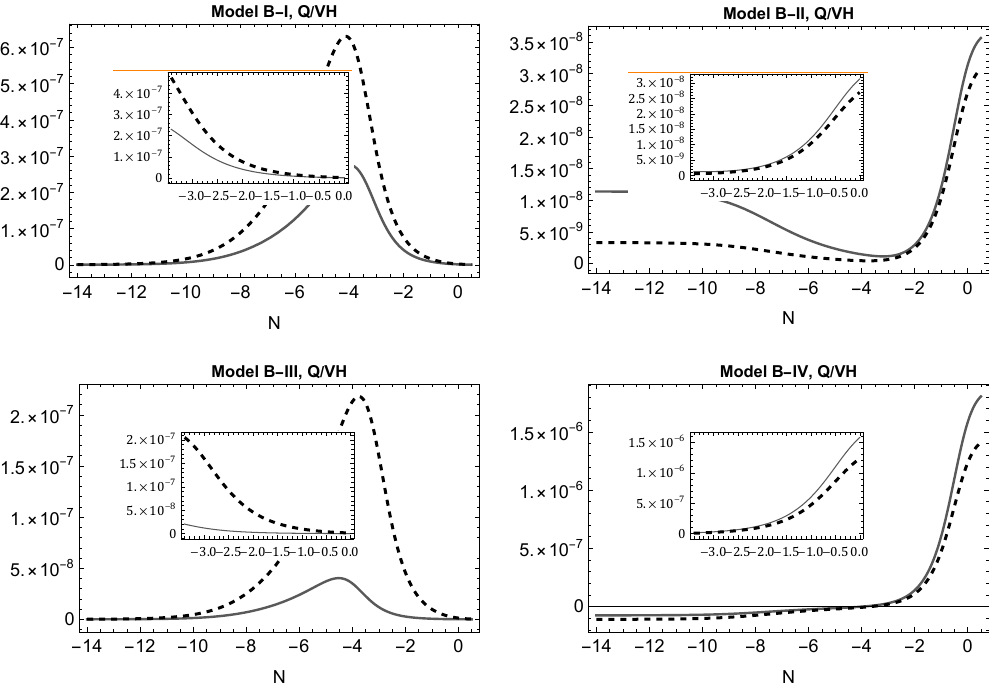}
		\caption{Evolution of the cosmological parameters for Model B, corresponding to the best-fit values obtained from MCMC. Solid and dashed lines represent the best-fit values for BASE+PP and BASE+DES datasets, respectively.}
		
		\label{fig:evo_modelB_sn}
		
	\end{figure*}

	\begin{figure*}
		\centering
		\includegraphics[scale=0.5]{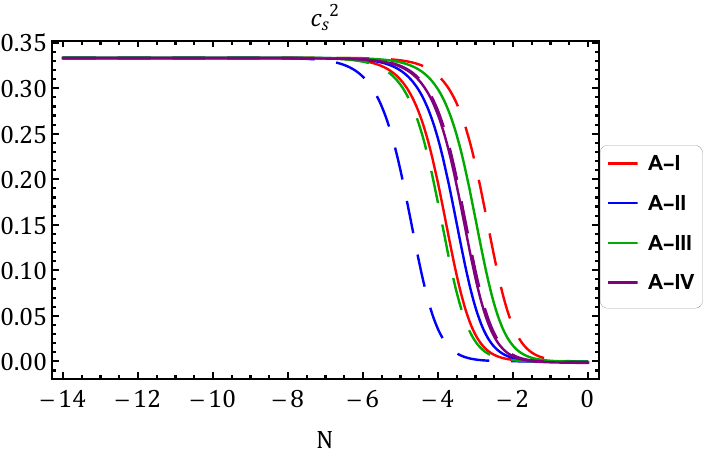}
		\hspace{0.2cm}
		\includegraphics[scale=0.5]{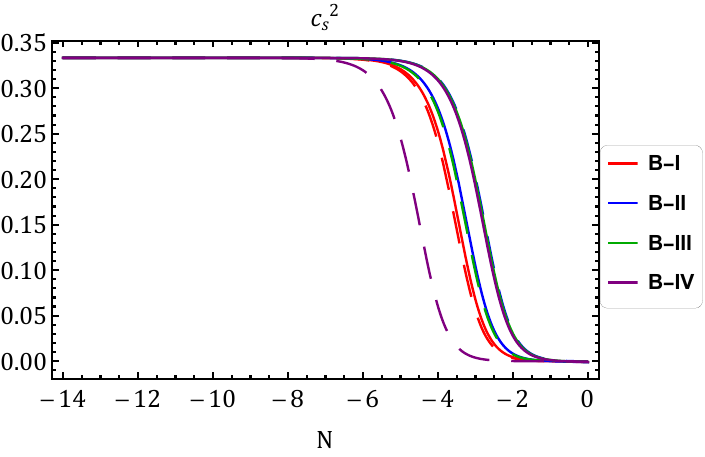}
		\caption{Evolution of the sound speed for Models A and B, corresponding to the best-fit values obtained from MCMC. Solid and dashed lines represent the best-fit values for BASE+PP and BASE+DES datasets, respectively.}
		
		\label{fig:sound_speed_sn}
		
	\end{figure*}
	
	\begin{table*}
		
		\adjustbox{angle=90, max height=\textheight}{
			\begin{tabular} { l  c c c c c c c c}
				\noalign{\vskip 3pt}\hline\noalign{\vskip 1.5pt}\hline\noalign{\vskip 5pt}
				\multicolumn{1}{c}{\bf Model A } &  \multicolumn{1}{c}{\bf I-BASE+PP} &  \multicolumn{1}{c}{\bf I-BASE+DES} &  \multicolumn{1}{c}{\bf II-BASE+PP} &  \multicolumn{1}{c}{\bf II-BASE+DES} &  \multicolumn{1}{c}{\bf III-BASE+PP} &  \multicolumn{1}{c}{\bf III-BASE+DES} &  \multicolumn{1}{c}{\bf IV-BASE+PP} &  \multicolumn{1}{c}{\bf IV-BASE+DES}\\
				\noalign{\vskip 3pt}\cline{2-9}\noalign{\vskip 3pt}
				
				Parameters &  68\% limits &  68\% limits &  68\% limits &  68\% limits &  68\% limits &  68\% limits &  68\% limits &  68\% limits\\
				\hline
				{\boldmath$\Omega_{\phi}  $} & $0.6910\pm 0.0027          $ & $0.6931^{+0.0027}_{-0.0022}$ & $0.6897\pm 0.0031          $ & $0.6933\pm 0.0025          $ & $0.6896\pm 0.0030          $ & $0.6932^{+0.0025}_{-0.0022}$ & $0.6887^{+0.0028}_{-0.0023}$ & $0.6919\pm 0.0025          $\\
				
				{\boldmath$H_0            $} & $67.08^{+0.62}_{-0.56}     $ & $69.63\pm 0.22             $ & $66.99^{+0.67}_{-0.57}     $ & $69.64\pm 0.22             $ & $66.96^{+0.65}_{-0.59}     $ & $69.61\pm 0.21             $ & $67.00^{+0.86}_{-0.63}     $ & $69.62\pm 0.21             $\\
				
				{\boldmath$x_0            $} & $1.1783^{+0.0027}_{-0.0039}$ & $1.1789^{+0.0024}_{-0.0027}$ & $1.1771^{+0.0032}_{-0.0040}$ & $1.1787\pm 0.0024          $ & $1.1777^{+0.0029}_{-0.0053}$ & $1.1789\pm 0.0024          $ & $1.1799\pm 0.0027          $ & $1.1801\pm 0.0023          $\\
				
				{\boldmath$\lambda        $} & $0.0080^{+0.0022}_{-0.0080}$ & $0.00446^{+0.00088}_{-0.0045}$ & $0.0078^{+0.0025}_{-0.0079}$ & $0.00339^{+0.00074}_{-0.0034}$ & $0.0096^{+0.0037}_{-0.0098}$ & $0.00425^{+0.00092}_{-0.0042}$ & $0.0053^{+0.0011}_{-0.0053}$ & $0.00371^{+0.00091}_{-0.0037}$\\
				
				{\boldmath$\theta         $} & $< 7.14\cdot 10^{-11}      $ & $< 1.17\cdot 10^{-10}      $ & $< 1.39\cdot 10^{-12}      $ & $< 1.49\cdot 10^{-12}      $ & $< 3.27\cdot 10^{-8}       $ & $< 1.29\cdot 10^{-7}       $ & $-0.0204^{+0.0049}_{-0.0033}$ & $-0.0115^{+0.0040}_{-0.0018}$\\
				
				{\boldmath$w_m            $} & $0.000173^{+0.000044}_{-0.00017}$ & $0.000174^{+0.000040}_{-0.00017}$ & $0.000177^{+0.000034}_{-0.00018}$ & $0.000160^{+0.000032}_{-0.00016}$ & $0.000167^{+0.000033}_{-0.00017}$ & $0.000158^{+0.000038}_{-0.00016}$ & $0.000337^{+0.000092}_{-0.00033}$ & $0.000254^{+0.000039}_{-0.00025}$\\
				
				{\boldmath$\Omega_r       $} & $\left(\,8.968\pm 0.057\,\right)\cdot 10^{-5}$ & $\left(\,8.941^{+0.047}_{-0.059}\,\right)\cdot 10^{-5}$ & $\left(\,8.993\pm 0.064\,\right)\cdot 10^{-5}$ & $\left(\,8.944^{+0.050}_{-0.056}\,\right)\cdot 10^{-5}$ & $\left(\,8.986\pm 0.064\,\right)\cdot 10^{-5}$ & $\left(\,8.939^{+0.046}_{-0.052}\,\right)\cdot 10^{-5}$ & $\left(\,8.920\pm 0.064\,\right)\cdot 10^{-5}$ & $\left(\,8.909\pm 0.057\,\right)\cdot 10^{-5}$\\
				
				{\boldmath$\Omega_{b}h^2  $} & $0.02233^{+0.00012}_{-0.00010}$ & $0.02267\pm 0.00011        $ & $0.02232^{+0.00012}_{-0.000094}$ & $0.02268\pm 0.00011        $ & $0.02233^{+0.00011}_{-0.000091}$ & $0.02268\pm 0.00011        $ & $0.02194^{+0.00020}_{-0.00013}$ & $0.02238^{+0.00017}_{-0.00013}$\\
				\hline
				\hline
				\noalign{\vskip 3pt}\cline{2-9}\noalign{\vskip 3pt}
				\multicolumn{1}{c}{\bf Model B } &  \multicolumn{1}{c}{\bf I-BASE+PP} &  \multicolumn{1}{c}{\bf I-BASE+DES} &  \multicolumn{1}{c}{\bf II-BASE+PP} &  \multicolumn{1}{c}{\bf II-BASE+DES} &  \multicolumn{1}{c}{\bf III-BASE+PP} &  \multicolumn{1}{c}{\bf III-BASE+DES} &  \multicolumn{1}{c}{\bf IV-BASE+PP} &  \multicolumn{1}{c}{\bf IV-BASE+DES}\\
				\noalign{\vskip 3pt}\cline{2-9}\noalign{\vskip 3pt}
				
				Parameters &  68\% limits &  68\% limits &  68\% limits &  68\% limits &  68\% limits &  68\% limits &  68\% limits &  68\% limits\\
				\hline
				{\boldmath$\Omega_{\phi}  $} & $0.6900^{+0.0031}_{-0.0028}$ & $0.6937^{+0.0026}_{-0.0021}$ & $0.6901\pm 0.0028          $ & $0.6936\pm 0.0022          $ & $0.6909^{+0.0030}_{-0.0023}$ & $0.6940^{+0.0022}_{-0.0019}$ & $0.6898^{+0.0033}_{-0.0028}$ & $0.6933^{+0.0029}_{-0.0022}$\\
				
				{\boldmath$H_0            $} & $67.03^{+0.64}_{-0.50}     $ & $69.65\pm 0.22             $ & $67.18^{+0.61}_{-0.53}     $ & $69.66\pm 0.22             $ & $67.13^{+0.65}_{-0.54}     $ & $69.65^{+0.24}_{-0.20}     $ & $67.00^{+0.67}_{-0.52}     $ & $69.65\pm 0.22             $\\
				
				{\boldmath$x_0            $} & $1.1780^{+0.0033}_{-0.0047}$ & $1.1795\pm 0.0026          $ & $1.1786^{+0.0028}_{-0.0054}$ & $1.1792^{+0.0021}_{-0.0025}$ & $1.1789^{+0.0031}_{-0.0044}$ & $1.1796^{+0.0020}_{-0.0025}$ & $1.1779^{+0.0030}_{-0.0044}$ & $1.1789^{+0.0027}_{-0.0024}$\\
				
				{\boldmath$\lambda        $} & $0.0097^{+0.0031}_{-0.0097}$ & $0.0048^{+0.0010}_{-0.0049}$ & $0.0113^{+0.0028}_{-0.011} $ & $0.00414^{+0.00068}_{-0.0041}$ & $0.00996^{+0.0026}_{-0.0099}$ & $0.00447^{+0.00069}_{-0.0045}$ & $0.0097^{+0.0022}_{-0.0098}$ & $0.00395^{+0.00086}_{-0.0040}$\\
				
				{\boldmath$\theta         $} & $< 2.25\cdot 10^{-9}       $ & $< 3.76\cdot 10^{-9}       $ & $< 2.25\cdot 10^{-8}       $ & $< 1.90\cdot 10^{-8}       $ & $< 6.61\cdot 10^{-7}       $ & $< 1.39\cdot 10^{-5}       $ & $\left(\,-1.05^{+1.6}_{+0.59}\,\right)\cdot 10^{-6}$ & $\left(\,-0.82^{+0.97}_{+0.62}\,\right)\cdot 10^{-6}$\\
				
				{\boldmath$w_m            $} & $0.000173^{+0.000039}_{-0.00017}$ & $0.000178^{+0.000046}_{-0.00018}$ & $0.000206^{+0.000055}_{-0.00021}$ & $0.000172^{+0.000035}_{-0.00017}$ & $0.000205^{+0.000042}_{-0.00021}$ & $0.000176^{+0.000058}_{-0.00018}$ & $0.000197^{+0.000052}_{-0.00020}$ & $0.000172^{+0.000042}_{-0.00017}$\\
				
				{\boldmath$\Omega_r       $} & $\left(\,8.986^{+0.061}_{-0.068}\,\right)\cdot 10^{-5}$ & $\left(\,8.933^{+0.049}_{-0.058}\,\right)\cdot 10^{-5}$ & $\left(\,8.979\pm 0.062\,\right)\cdot 10^{-5}$ & $\left(\,8.929\pm 0.050\,\right)\cdot 10^{-5}$ & $\left(\,8.965\pm 0.059\,\right)\cdot 10^{-5}$ & $\left(\,8.920\pm 0.049\,\right)\cdot 10^{-5}$ & $\left(\,8.984\pm 0.063\,\right)\cdot 10^{-5}$ & $\left(\,8.939^{+0.047}_{-0.060}\,\right)\cdot 10^{-5}$\\
				
				{\boldmath$\Omega_{b}h^2  $} & $0.02233^{+0.00011}_{-0.000094}$ & $0.02269\pm 0.00011        $ & $0.02234^{+0.00011}_{-0.000093}$ & $0.02268\pm 0.00010        $ & $0.02232^{+0.00012}_{-0.000089}$ & $0.02268\pm 0.00010        $ & $0.02231^{+0.00012}_{-0.000091}$ & $0.02269\pm 0.00011        $\\
				
				\noalign{\vskip 3pt}\hline\noalign{\vskip 1.5pt}\hline\noalign{\vskip 5pt}
				\multicolumn{1}{c}{\bf } &  \multicolumn{1}{c}{\bf A-1-BASE+HCW} &  \multicolumn{1}{c}{\bf A-2-BASE+HCW} &  \multicolumn{1}{c}{\bf A-3-BASE+HCW} &  \multicolumn{1}{c}{\bf A-4-BASE+HCW} &  \multicolumn{1}{c}{\bf B-1-BASE+HCW} &  \multicolumn{1}{c}{\bf B-2-BASE+HCW} &  \multicolumn{1}{c}{\bf B-3-BASE+HCW} &  \multicolumn{1}{c}{\bf B-4-BASE+HCW}\\
				\noalign{\vskip 3pt}\cline{2-9}\noalign{\vskip 3pt}
				
				Parameter &  68\% limits &  68\% limits &  68\% limits &  68\% limits &  68\% limits &  68\% limits &  68\% limits &  68\% limits\\
				\hline
				{\boldmath$\Omega_{\phi}  $} & $0.6927\pm 0.0025          $ & $0.6916^{+0.0030}_{-0.0026}$ & $0.6917^{+0.0029}_{-0.0032}$ & $0.6914\pm 0.0025          $ & $0.6918^{+0.0026}_{-0.0022}$ & $0.6918\pm 0.0029          $ & $0.6934^{+0.0028}_{-0.0019}$ & $0.6923^{+0.0029}_{-0.0025}$\\
				
				{\boldmath$H_0            $} & $67.87^{+0.55}_{-0.80}     $ & $67.68^{+0.62}_{-0.74}     $ & $67.71^{+0.57}_{-0.82}     $ & $67.77\pm 0.63             $ & $67.64^{+0.53}_{-0.69}     $ & $67.71^{+0.56}_{-0.67}     $ & $67.96^{+0.64}_{-0.74}     $ & $67.79^{+0.59}_{-0.72}     $\\
				
				{\boldmath$x_0            $} & $1.1802^{+0.0025}_{-0.0046}$ & $1.1795^{+0.0034}_{-0.0045}$ & $1.1793^{+0.0031}_{-0.0047}$ & $1.1820^{+0.0022}_{-0.0028}$ & $1.1803^{+0.0028}_{-0.0046}$ & $1.1799^{+0.0033}_{-0.0053}$ & $1.1815^{+0.0027}_{-0.0047}$ & $1.1816^{+0.0052}_{-0.0047}$\\
				
				{\boldmath$\lambda        $} & $0.0091^{+0.0025}_{-0.0091}$ & $0.0101^{+0.0032}_{-0.0099}$ & $0.0091^{+0.0050}_{-0.0091}$ & $0.0061^{+0.0017}_{-0.0061}$ & $0.0118^{+0.0026}_{-0.012} $ & $0.0108^{+0.0037}_{-0.011} $ & $0.0115^{+0.0033}_{-0.012} $ & $0.0142^{+0.0099}_{-0.014} $\\
				
				{\boldmath$\theta         $} & $< 1.04\cdot 10^{-8}       $ & $\left(\,0.610^{-0.031}_{-0.55}\,\right)\cdot 10^{-10}$ & $< 3.75\cdot 10^{-6}       $ & $-0.0181^{+0.0054}_{-0.0028}$ & $< 3.32\cdot 10^{-7}       $ & $< 5.95\cdot 10^{-7}       $ & $< 0.000133                $ & $> -1.90\cdot 10^{-8}      $\\
				
				{\boldmath$w_m            $} & $0.000149^{+0.000042}_{-0.00015}$ & $0.000163^{+0.000054}_{-0.00016}$ & $0.000150^{+0.000028}_{-0.00015}$ & $0.000244^{+0.000053}_{-0.00024}$ & $0.000176^{+0.000055}_{-0.00017}$ & $0.000166^{+0.000038}_{-0.00017}$ & $0.000137^{+0.000025}_{-0.00014}$ & $0.000170^{+0.000036}_{-0.00017}$\\
				
				{\boldmath$\Omega_r       $} & $\left(\,8.940\pm 0.057\,\right)\cdot 10^{-5}$ & $\left(\,8.975\pm 0.063\,\right)\cdot 10^{-5}$ & $\left(\,8.960\pm 0.063\,\right)\cdot 10^{-5}$ & $\left(\,8.899\pm 0.054\,\right)\cdot 10^{-5}$ & $\left(\,8.953^{+0.049}_{-0.059}\,\right)\cdot 10^{-5}$ & $\left(\,8.950^{+0.059}_{-0.068}\,\right)\cdot 10^{-5}$ & $\left(\,8.937\pm 0.054\,\right)\cdot 10^{-5}$ & $\left(\,8.945\pm 0.062\,\right)\cdot 10^{-5}$\\
				
				{\boldmath$\Omega_{b}h^2  $} & $0.022410^{+0.000084}_{-0.00015}$ & $0.022383^{+0.000091}_{-0.00013}$ & $0.022394^{+0.000090}_{-0.00015}$ & $0.02202^{+0.00016}_{-0.00012}$ & $0.022389^{+0.000095}_{-0.00013}$ & $0.022384^{+0.000096}_{-0.00013}$ & $0.022404^{+0.000091}_{-0.00016}$ & $0.022386^{+0.000096}_{-0.00013}$\\
				\hline
				
				\noalign{\vskip 3pt}\hline\noalign{\vskip 1.5pt}\hline\noalign{\vskip 5pt}
				\multicolumn{1}{c}{\bf Models} &  \multicolumn{1}{c}{\bf A-1-CDB+HCW} &  \multicolumn{1}{c}{\bf A-2-CDB+HCW} &  \multicolumn{1}{c}{\bf A-3-CDB+HCW} &  \multicolumn{1}{c}{\bf A-4-CDB+HCW} &  \multicolumn{1}{c}{\bf B-1-CDB+HCW} &  \multicolumn{1}{c}{\bf B-2-CDB+HCW} &  \multicolumn{1}{c}{\bf B-3-CDB+HCW} &  \multicolumn{1}{c}{\bf B-4-CDB+HCW}\\
				\noalign{\vskip 3pt}\cline{2-9}\noalign{\vskip 3pt}
				
				Parameter &  68\% limits &  68\% limits &  68\% limits &  68\% limits &  68\% limits &  68\% limits &  68\% limits &  68\% limits\\
				\hline
				{\boldmath$\Omega_{\phi}  $} & $0.7027^{+0.0083}_{-0.0060}$ & $0.7017^{+0.0067}_{-0.0076}$ & $0.6992\pm 0.0077          $ & $0.6823^{+0.0097}_{-0.012} $ & $0.7023^{+0.0065}_{-0.0075}$ & $0.6990^{+0.0087}_{-0.0073}$ & $0.7039\pm 0.0081          $ & $0.7012^{+0.0094}_{-0.0078}$\\
				
				{\boldmath$H_0            $} & $68.5\pm 1.3               $ & $69.0^{+1.3}_{-1.5}        $ & $67.7^{+1.4}_{-1.3}        $ & $66.74^{+0.54}_{-1.1}      $ & $68.4^{+1.4}_{-1.7}        $ & $68.4^{+1.4}_{-1.1}        $ & $68.9^{+1.3}_{-1.2}        $ & $68.5^{+1.3}_{-1.1}        $\\
				
				{\boldmath$x_0            $} & $1.1899\pm 0.0074          $ & $1.1879^{+0.0051}_{-0.0080}$ & $1.1864^{+0.0065}_{-0.0075}$ & $1.1857\pm 0.0095          $ & $1.1890^{+0.0057}_{-0.0069}$ & $1.1850^{+0.0074}_{-0.0065}$ & $1.1903\pm 0.0076          $ & $1.1877^{+0.0085}_{-0.0073}$\\
				
				{\boldmath$\lambda        $} & $0.0127^{+0.0064}_{-0.013} $ & $0.0093^{+0.0026}_{-0.0089}$ & $0.0111^{+0.0027}_{-0.011} $ & $0.0158^{+0.0031}_{-0.015} $ & $0.0110^{+0.0027}_{-0.011} $ & $0.0079^{+0.0015}_{-0.0079}$ & $0.0111^{+0.0040}_{-0.011} $ & $0.0100^{+0.0023}_{-0.010} $\\
				
				{\boldmath$\theta         $} & $< 7.45\cdot 10^{-9}       $ & $\left(\,0.555^{-0.019}_{-0.48}\,\right)\cdot 10^{-10}$ & $< 2.12\cdot 10^{-6}       $ & $-0.056^{+0.031}_{-0.022}  $ & $< 3.65\cdot 10^{-6}       $ & $< 1.39\cdot 10^{-7}       $ & $< 1.42\cdot 10^{-5}       $ & $> -3.85\cdot 10^{-12}     $\\
				
				{\boldmath$w_m            $} & $0.00038^{+0.00021}_{-0.00038}$ & $0.000297^{+0.000055}_{-0.00030}$ & $0.00040^{+0.00011}_{-0.00040}$ & $0.00037^{+0.00062}_{-0.00037}$ & $0.000306^{+0.000043}_{-0.00030}$ & $0.00039^{+0.00010}_{-0.00039}$ & $0.00045^{+0.00013}_{-0.00045}$ & $0.000412^{+0.000087}_{-0.00041}$\\
				
				{\boldmath$\Omega_r       $} & $\left(\,8.895\pm 0.089\,\right)\cdot 10^{-5}$ & $\left(\,8.885^{+0.097}_{-0.087}\,\right)\cdot 10^{-5}$ & $\left(\,8.917\pm 0.087\,\right)\cdot 10^{-5}$ & $\left(\,8.95^{+0.12}_{-0.090}\,\right)\cdot 10^{-5}$ & $\left(\,8.90^{+0.10}_{-0.077}\,\right)\cdot 10^{-5}$ & $\left(\,8.887\pm 0.078\,\right)\cdot 10^{-5}$ & $\left(\,8.865\pm 0.085\,\right)\cdot 10^{-5}$ & $\left(\,8.885\pm 0.082\,\right)\cdot 10^{-5}$\\
				
				{\boldmath$\Omega_{b}h^2  $} & $0.02160^{+0.00050}_{-0.00041}$ & $0.02184\pm 0.00050        $ & $0.02138^{+0.00046}_{-0.00042}$ & $0.02110^{+0.00031}_{-0.00041}$ & $0.02167^{+0.00044}_{-0.00037}$ & $0.02166^{+0.00054}_{-0.00040}$ & $0.02179^{+0.00049}_{-0.00042}$ & $0.02166^{+0.00049}_{-0.00040}$\\
				
				\hline
				\hline
			\end{tabular}
		}
		\caption{The best fit values of the model parameters at 68 \% level. Here, BASE and CDB refers to CC+DBAO+PLA, and CC+DBAO+BBN data sets respectively. } 
		\label{tab:model_A+B_best_fit_sn}
	\end{table*}
	
	\begin{table*}
		\centering
		\scriptsize
		\begin{tabular}{lcccccccccccc}
			\hline \hline
			& \multicolumn{2}{c}{Model A-I} & \multicolumn{2}{c}{Model A-II} & \multicolumn{2}{c}{Model A-III} & \multicolumn{2}{c}{Model A-IV} & \multicolumn{2}{c}{Model B-I} & \multicolumn{2}{c}{Model B-II} \\
			& BASE+PP & BASE+DES & BASE+PP & BASE+DES & BASE+PP & BASE+DES & BASE+PP & BASE+DES & BASE+PP & BASE+DES & BASE+PP & BASE+DES \\
			\hline
			$\chi^2$ & 1441.07 & 1727.17 & 1441.05 & 1726.33 & 1440.75 & 1726.99 & 1435.20 & 1725.28 & 1441.21 & 1726.49 & 1441.27 & 1726.44 \\
			AIC      & 1457.07 & 1743.17 & 1457.05 & 1742.33 & 1456.75 & 1742.99 & 1451.20 & 1741.28 & 1457.21 & 1742.49 & 1457.27 & 1742.44 \\
			BIC      & 1500.25 & 1787.45 & 1500.23 & 1786.60 & 1499.93 & 1787.27 & 1494.38 & 1785.55 & 1500.39 & 1786.76 & 1500.45 & 1786.72 \\
			\hline
			& \multicolumn{2}{c}{Model B-III} & \multicolumn{2}{c}{Model B-IV} & \multicolumn{2}{c}{\(\Lambda\)CDM} & & & & & & \\
			& BASE+PP & BASE+DES & BASE+PP & BASE+DES & BASE+PP & BASE+DES & & & & & & \\
			\hline
			$\chi^2$ & 1441.09 & 1726.48 & 1441.11 & 1726.78 & 1441.55 & 1725.23 &&&&&&\\
			AIC      & 1457.09 & 1742.48 & 1457.11 & 1742.78 & 1449.55 & 1733.23 &&&&&&\\
			BIC      & 1500.27 & 1786.76 & 1500.29 & 1787.05 & 1471.14 & 1755.37  &&&&&&\\
			\hline \hline
			& \multicolumn{12}{c}{\bf \boldmath $\Delta \chi^2$ for {H0LiCOW Data}} \\
			\hline
			& A-I & A-II & A-III & A-IV & B-I & B-II & B-III & B-IV & \(\Lambda\)CDM & & & \\
			\hline
			{\scriptsize BASE+HCW }     & 1.0 & 1.58 & 1.35 & -6.13 & 0.87 & 1.9 & 0.87 & 0.62 & 0 & & & \\
			{\scriptsize CC+DBAO+BBN+HCW } & -0.79 & -1.29& 0.28 & 0.02 & -0.84 & -0.62 & -1.21 & -0.44 & 0 & & & \\
			{\scriptsize BASE+BBN+HCW }   & -13.08 & -12.7 & -12.98 & -18.81 & -13.33 & -13.88 & -12.48 & -13.39 & 0 & & & \\
			\hline \hline
		\end{tabular}
		\caption{Comparison of $\chi^2$, AIC, and BIC values for different models and $\Lambda$CDM across various dataset combinations. Here, $\Delta \chi^2 \equiv \chi^2_{\rm Model} - \chi^2_{\rm \Lambda CDM}$.}
		
		\label{tab:stat_aic_bic}
	\end{table*}

	\section{Some Additional Results with Strong Lensing Data \label{sec:holicow_discussion}}
	
	In this section, we employ an independent observational probe to investigate whether the considered models can alleviate the Hubble tension. Type Ia Supernovae provide excellent constraints on the Hubble constant. Previously, we used two different supernova catalogs and found that both interacting models fit the data remarkably well. Both Models A and B show good fits, with improved chi-squared values, and produce stable de-Sitter solutions, yielding \(H_0 \approx 67.3\) km/s/Mpc for the PP data and \(H_0 \approx 69.6\) km/s/Mpc for the DES data. This raises a natural question: how much upward shift in \(H_0\) can be realized within the considered models? An independent probe of \(H_0\) may help answer this.
	
	Gravitational lensing provides such an independent measurement of the Hubble constant \cite{H0LiCOW:2019pvv,Hu:2023jqc,H0LiCOW:2016xpx}. We use observational samples from six distinct lenses—B1608+656, RXJ1131-1231, SDSS 1206+4332, WFI2033-4723, HE0435-1223, and PG1115+080--from the H0LiCOW (H0 Lenses in COSMOGRAIL's Wellspring) program \cite{H0LiCOW:2019pvv}. The likelihood estimation relies on several quantities, including the time-delay distance \(D_{\Delta t}\), which is inferred from observed time delays \(\Delta t_{\rm obs}\). This quantity itself depends on combinations of angular diameter distances, given by
	\begin{equation}
		D_{\Delta t} = (1+z_{\rm d}) \frac{D_d D_s}{D_{d s}} \ ,
	\end{equation} 
	where \(z_{\rm d}\) and \(z_s\) are the lens and source redshifts, respectively, and \(c\) is the speed of light in km/s. \(D_{d}\) and \(D_{s}\) are the angular diameter distances to the lens and source, respectively, while \(D_{ds}\) is the angular diameter distance between the source and the lens. The angular diameter distances to the lens and source are given by
	\begin{equation}
		D_{d,s} = \frac{c}{(1+z_{d,s})} \int_{0}^{z_{d,s}} \frac{1}{H(z)} dz \ ,
	\end{equation}
	and the distance between the source and lens for a flat metric is determined as \cite{Suyu:2018vqs}:
	\begin{equation}
		D_{d s} = \frac{c}{(1+z_{s})} \int_{z_{d}}^{z_{s}} \frac{1}{H(z)} dz \ .
	\end{equation}
	The likelihood estimation is obtained by computing these distances at the corresponding redshifts. The observational samples and the likelihood estimation methodology, including the Python implementation, can be found at this \href{https://github.com/shsuyu/H0LiCOW-public/tree/master/MontePython_cosmo_sampling/likelihoods/timedelay_6lenses}{link}\footnote{The H0LiCOW GitHub repository is available at \url{https://github.com/shsuyu/H0LiCOW-public}} \cite{Suyu:2009by,Suyu:2013kha,H0LiCOW:2016qrm,H0LiCOW:2018tyj,H0LiCOW:2019xdh,Jee:2019hah,H0LiCOW:2019mdu}. 

	The flat $\Lambda$CDM model for all the considered lenses (H0LiCOW only) yields \(H_0 = 73.3^{+1.9}_{-1.6}\) km/s/Mpc \cite{H0LiCOW:2019pvv}. Nevertheless, this observation alone cannot constrain other parameters, such as the dark energy density \(\Omega_{\Lambda}\). Therefore, we include an external probe along with this dataset to obtain consistent constraints on the remaining parameters. As external probes, we consider the following three combinations of datasets:
	\begin{equation}
		\begin{aligned}
			\rm &(i) \ \text{CC+BAO+PLA+HCW,} \\ 
			\rm &(ii) \ \text{CC+BAO+BBN+HCW,}\\ 
			\rm &(iii)\  \text{CC+BAO+PLA+BBN+HCW}\ . \nonumber
		\end{aligned}
	\end{equation}
	In the second combination, we exclude the Planck data, as it tends to drive \(H_0\) towards \(\sim 67.0\) km/s/Mpc. However, BAO data alone cannot constrain \(H_0\) or \(\Omega_{b} h^2\). Therefore, BBN data is included to constrain the baryon density. Finally, we incorporate all the datasets, which allows for stringent bounds on the model parameters and aids in distinguishing the behavior of the different models.
	
	The best-fit values of the model parameters are reported in Tabs. \ref{tab:model_A+B_best_fit_sn} and \ref{tab:lcdm_holicow}, and the corresponding posterior distributions are shown in Fig. \ref{fig:holicow_chains}. The flat \(\Lambda\)CDM model, with free parameters \((\Omega_{\Lambda}, H_0, \Omega_{r}, \Omega_{b}h^2)\), is fitted against these datasets, yielding \(H_0 = 67.71\) km/s/Mpc for combinations (i) and (ii), while the third combination yields \(H_0 = 68.79\) km/s/Mpc. 
	
	The submodels of A and B yield nearly \(H_0 \sim 67.72\) km/s/Mpc for the BASE+HCW dataset, with dark energy density approximately \(69.1\%\). The dataset produces a similar magnitude of \(w_m\) as obtained for the previous samples. For the CDB+HCW dataset, the models generally show an enhanced \(H_0\) value around \(68-69\) km/s/Mpc, with dark energy density close to \(70\%\), whereas Model A-IV yields a slightly lower \(H_0\), nearly \(66\) km/s/Mpc, with \(68\%\) dark energy density. For this dataset, a higher magnitude of \(w_m \sim 3.0 \cdot 10^{-4}\) is recorded.

	Considering the third combination of datasets, nearly all models yield \(H_0 \sim 68.8\) km/s/Mpc, consistent with the value obtained from the fiducial model. However, Model A-IV shows a mildly lower \(H_0 \sim 67.95\) km/s/Mpc. The models yield approximately \(69.3\%\) dark energy density. 
	
	The evolution of the cosmological parameters, including the interaction parameter, for all these combinations of datasets is illustrated in Figs. \ref{fig:model_a_holic_evo} and \ref{fig:model_b_holic_evo}. The behavior of the model parameters follows trends similar to those obtained for the supernova datasets. {The sound speed is plotted in Fig.~\ref{fig:sound_speed_hcw} for each dataset. The evolution shows that the sound speed remains positive in the deep past and becomes extremely small at the present and in the late-time future, providing a necessary consistency check related to the absence of ghost-like behavior.
	 } 
	
	The chi-squared differences of the models relative to \(\Lambda\)CDM are outlined in Tab. \ref{tab:stat_aic_bic}. For the BASE+HCW datasets, nearly all models yield slightly positive values, except for Model A-IV, which gives a higher negative value, indicating an excellent fit. For CDB+HCW and BASE+BBN+HCW datasets, nearly all models yield negative differences; for the latter dataset, all interacting models exhibit comparatively lower chi-squared values.

	\begin{figure*}
		\includegraphics[scale=0.45]{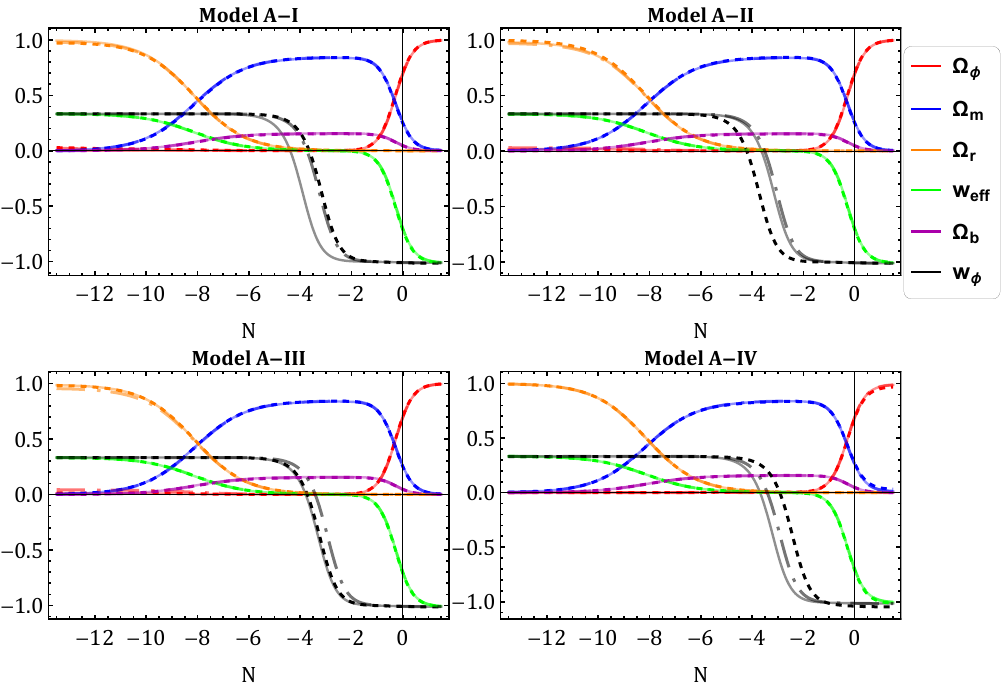}
		\includegraphics[scale=0.45]{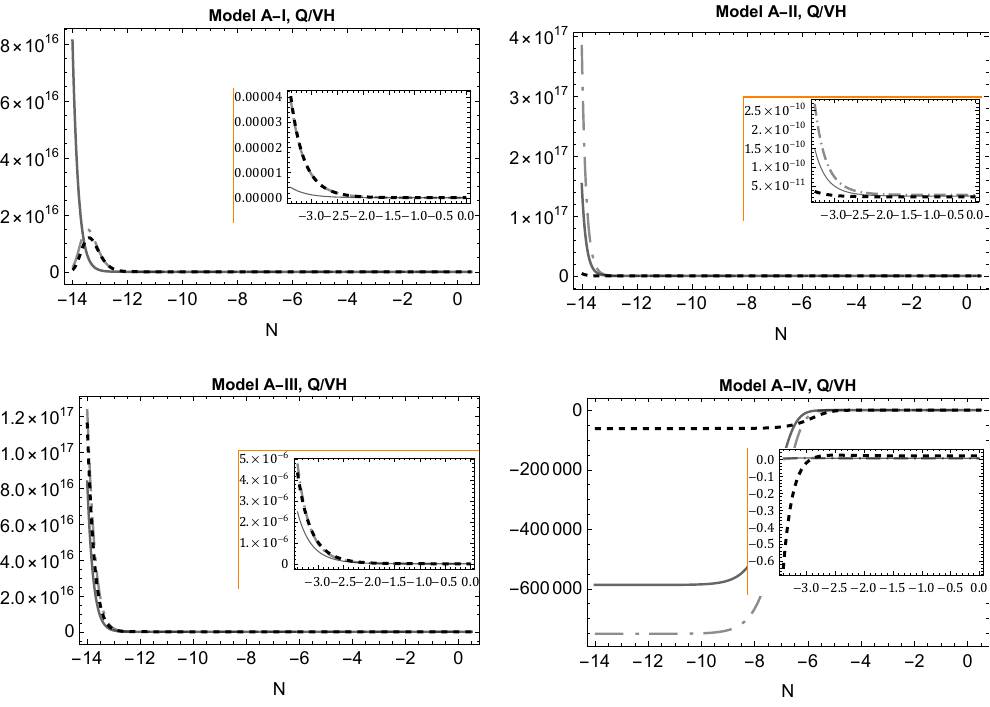}
		\caption{Evolution of cosmological parameters of Model A for the best-fit values obtained using H0LiCOW data. The solid, dashed, and dot-dashed lines correspond to the BASE+HCW, CDB+HCW, and BASE+BBN+HCW datasets, respectively.}
		
		\label{fig:model_a_holic_evo}
	\end{figure*}
	
	\begin{figure*}
		\includegraphics[scale=0.45]{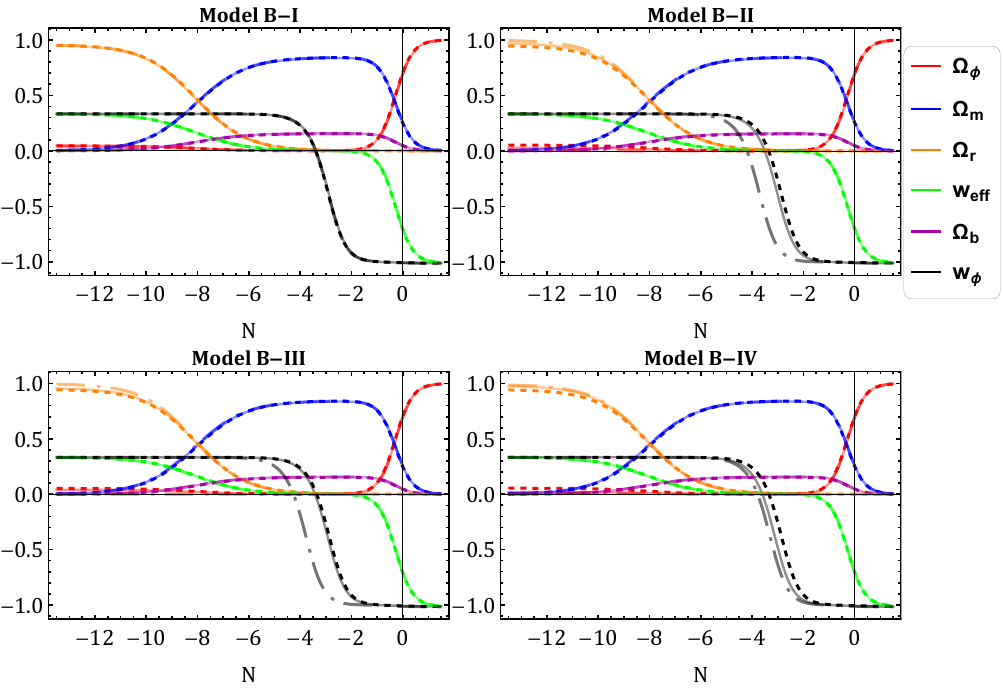}
		\includegraphics[scale=0.45]{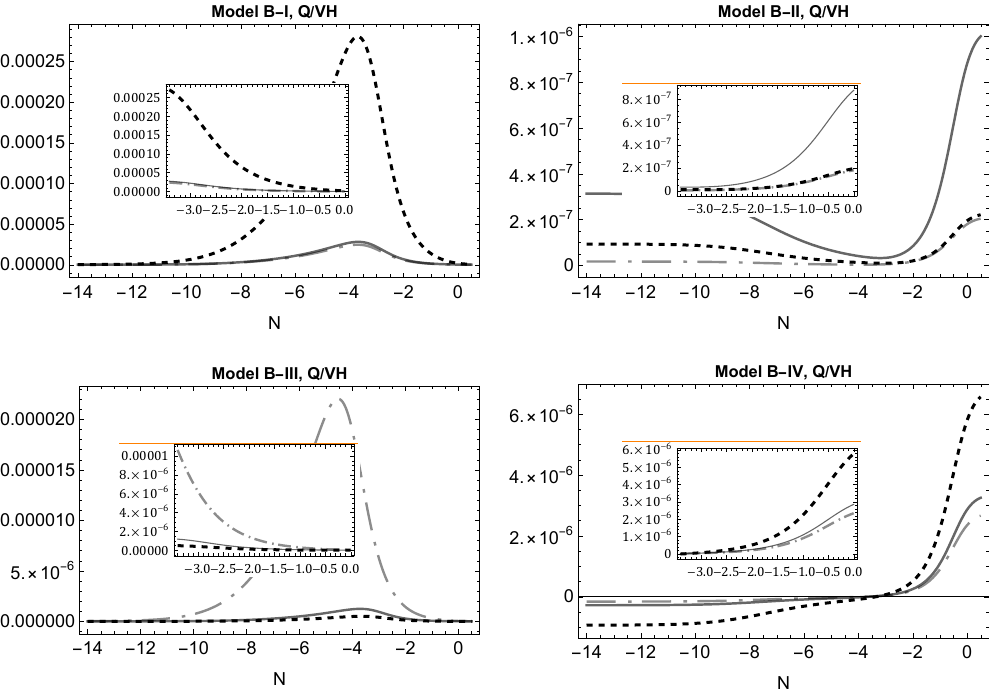}
		\caption{Evolution of cosmological parameters of Model B for the best-fit values obtained using H0LiCOW data. The solid, dashed, and dot-dashed lines correspond to the BASE+HCW, CDB+HCW, and BASE+BBN+HCW datasets, respectively.}
		\label{fig:model_b_holic_evo}
	\end{figure*}

	\begin{figure*}
		\includegraphics[scale=0.5]{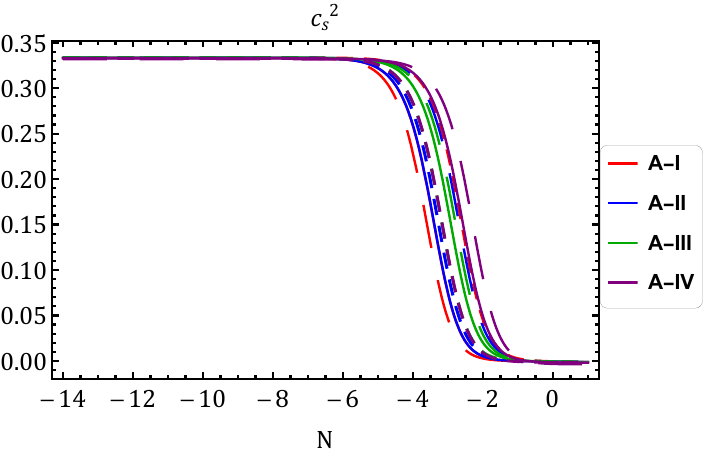}
		\includegraphics[scale=0.5]{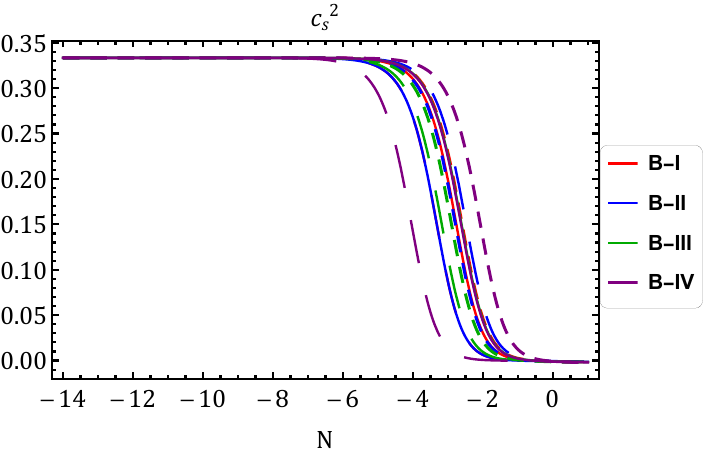}
			\caption{Evolution of sound speed of Models A and B for the best-fit values. The solid, long-dashed, and dashed lines correspond to the BASE+HCW, CDB+HCW, and BASE+BBN+HCW datasets, respectively.}
			\label{fig:sound_speed_hcw}
	\end{figure*}

	\begin{table*}[t]
		\adjustbox{angle=90,  height=\textheight}{
			\begin{tabular} { l  c c c c c c c c}
				\noalign{\vskip 3pt}\hline\noalign{\vskip 1.5pt}\hline\noalign{\vskip 5pt}
				\multicolumn{1}{c}{\bf Models } &  \multicolumn{1}{c}{\bf A-1-BASE+BBN+HCW} &  \multicolumn{1}{c}{\bf A-2-BASE+BBN+HCW} &  \multicolumn{1}{c}{\bf A-3-BASE+BBN+HCW} &  \multicolumn{1}{c}{\bf A-4-BASE+BBN+HCW} &  \multicolumn{1}{c}{\bf B-1-BASE+BBN+HCW} &  \multicolumn{1}{c}{\bf B-2-BASE+BBN+HCW} &  \multicolumn{1}{c}{\bf B-3-BASE+BBN+HCW} &  \multicolumn{1}{c}{\bf B-4-BASE+BBN+HCW}\\
				\noalign{\vskip 3pt}\cline{2-9}\noalign{\vskip 3pt}
				
				Parameter &  68\% limits &  68\% limits &  68\% limits &  68\% limits &  68\% limits &  68\% limits &  68\% limits &  68\% limits\\
				\hline
				{\boldmath$\Omega_{\phi}  $} & $0.6932\pm 0.0027          $ & $0.6938^{+0.0028}_{-0.0023}$ & $0.6931\pm 0.0029          $ & $0.6902^{+0.0027}_{-0.0019}$ & $0.6939^{+0.0025}_{-0.0017}$ & $0.6946\pm 0.0026          $ & $0.6939^{+0.0028}_{-0.0022}$ & $0.6947^{+0.0035}_{-0.0020}$\\
				
				{\boldmath$H_0            $} & $68.79\pm 0.70             $ & $68.72^{+0.65}_{-0.75}     $ & $68.75\pm 0.69             $ & $67.95^{+0.62}_{-0.46}     $ & $68.80^{+0.62}_{-0.77}     $ & $68.98\pm 0.69             $ & $68.70^{+0.60}_{-0.68}     $ & $68.84\pm 0.63             $\\
				
				{\boldmath$x_0            $} & $1.1810^{+0.0032}_{-0.0044}$ & $1.1815^{+0.0031}_{-0.0042}$ & $1.1807^{+0.0034}_{-0.0043}$ & $1.1810^{+0.0026}_{-0.0034}$ & $1.1819^{+0.0029}_{-0.0041}$ & $1.1821^{+0.0030}_{-0.0041}$ & $1.1833^{+0.0035}_{-0.0069}$ & $1.1844^{+0.0083}_{-0.0066}$\\
				
				{\boldmath$\lambda        $} & $0.0103^{+0.0030}_{-0.010} $ & $0.0105^{+0.0029}_{-0.010} $ & $0.00998^{+0.0026}_{-0.010}$ & $0.0077^{+0.0020}_{-0.0077}$ & $0.0112^{+0.0050}_{-0.011} $ & $0.0100^{+0.0030}_{-0.010} $ & $0.015^{+0.019}_{-0.016}   $ & $0.016\pm 0.011            $\\
				
				{\boldmath$\theta         $} & $< 7.42\cdot 10^{-9}       $ & $\left(\,0.698^{-0.085}_{-0.64}\,\right)\cdot 10^{-10}$ & $< 5.39\cdot 10^{-6}       $ & $-0.0160^{+0.0051}_{-0.0039}$ & $< 2.88\cdot 10^{-7}       $ & $< 1.23\cdot 10^{-7}       $ & $< 0.000686                $ & $> -1.56\cdot 10^{-8}      $\\
				
				{\boldmath$w_m            $} & $0.000132^{+0.000032}_{-0.00013}$ & $0.000120^{+0.000039}_{-0.00012}$ & $0.000126^{+0.000028}_{-0.00013}$ & $0.000241^{+0.000075}_{-0.00024}$ & $0.000129^{+0.000035}_{-0.00013}$ & $0.000134^{+0.000028}_{-0.00014}$ & $0.000115^{+0.000024}_{-0.00011}$ & $0.000109^{+0.000023}_{-0.00011}$\\
				
				{\boldmath$\Omega_r       $} & $\left(\,8.956\pm 0.059\,\right)\cdot 10^{-5}$ & $\left(\,8.959\pm 0.058\,\right)\cdot 10^{-5}$ & $\left(\,8.964^{+0.056}_{-0.067}\,\right)\cdot 10^{-5}$ & $\left(\,8.940\pm 0.052\,\right)\cdot 10^{-5}$ & $\left(\,8.948^{+0.049}_{-0.059}\,\right)\cdot 10^{-5}$ & $\left(\,8.938^{+0.050}_{-0.063}\,\right)\cdot 10^{-5}$ & $\left(\,8.945\pm 0.057\,\right)\cdot 10^{-5}$ & $\left(\,8.928^{+0.050}_{-0.071}\,\right)\cdot 10^{-5}$\\
				
				{\boldmath$\Omega_{b}h^2  $} & $0.02244^{+0.00014}_{-0.00020}$ & $0.02242^{+0.00014}_{-0.00019}$ & $0.02243^{+0.00017}_{-0.00020}$ & $0.02203^{+0.00018}_{-0.00012}$ & $0.02244^{+0.00014}_{-0.00020}$ & $0.02245\pm 0.00016        $ & $0.02241^{+0.00013}_{-0.00018}$ & $0.02242^{+0.00013}_{-0.00018}$\\
				
				\noalign{\vskip 3pt}\hline\noalign{\vskip 1.5pt}\hline\noalign{\vskip 5pt}
				\multicolumn{1}{c}{\boldmath \bf $\Lambda$CDM } &  \multicolumn{1}{c}{\bf BASE+PP} &  \multicolumn{1}{c}{\bf BASE+DES} &  \multicolumn{1}{c}{\bf BASE+HCW} &  \multicolumn{1}{c}{\bf CC+DBAO+BBN+HCW} &  \multicolumn{1}{c}{\bf BASE+BBN+HCW} &  \multicolumn{1}{c}{\bf } &  \multicolumn{1}{c}{\bf } &  \multicolumn{1}{c}{\bf }\\
				\noalign{\vskip 3pt}\cline{2-9}\noalign{\vskip 3pt}
				
				Parameter &  68\% limits &  68\% limits &  68\% limits &  68\% limits &  68\% limits &  68\% limits &  68\% limits &  68\% limits\\
				\hline
				{\boldmath$ \Omega_{\Lambda}$} & $0.6806 \pm 0.00204$ & $0.6891 \pm 0.0020$ & $0.6843^{+0.0022}_{-0.0032}$& $0.7038\pm 0.0082$& $0.6869^{+0.0026}_{-0.0030}$&&&\\
				
				{\boldmath$ H_0$} & $66.1^{+1.2}_{-1.4}$ & $69.38\pm 0.20$ & $67.72^{0.90}_{-0.64}$&$67.7\pm1.3$ & $68.79^{+0.51}_{-0.37}$&&&\\

				{\boldmath$ \Omega_{b}h^2$} & $0.02221^{+0.00020}_{-0.00014} $ & $ 0.022495^{+0.0000620}_{-0.000010}$ & $0.02237^{+0.00014}_{-0.00010}$& $0.02121\pm 0.00046$& $0.02234^{+0.00012}_{-0.000084}$&&&\\
				
				{\boldmath$ \Omega_{r}$} & $(8.769 \pm 0.063) \cdot 10^{-5} $ & $ (8.661\pm 0.045) \cdot 10^{-5}$ & $(8.724^{+0.066}_{-0.055}) \cdot 10^{-5}$& $(8.815 \pm 0.083) \cdot 10^{-5}$& $(8.718^{+0.057}_{-0.048}) \cdot 10^{-5}$&&&\\
				
				\hline
				\hline
			\end{tabular}
		}
		\caption{The best fit value for the model parameters at 68\% CL  corresponding to the specified datasets.}
		
		\label{tab:lcdm_holicow}
	\end{table*}
	
	\begin{figure*}
		\includegraphics[scale=0.4]{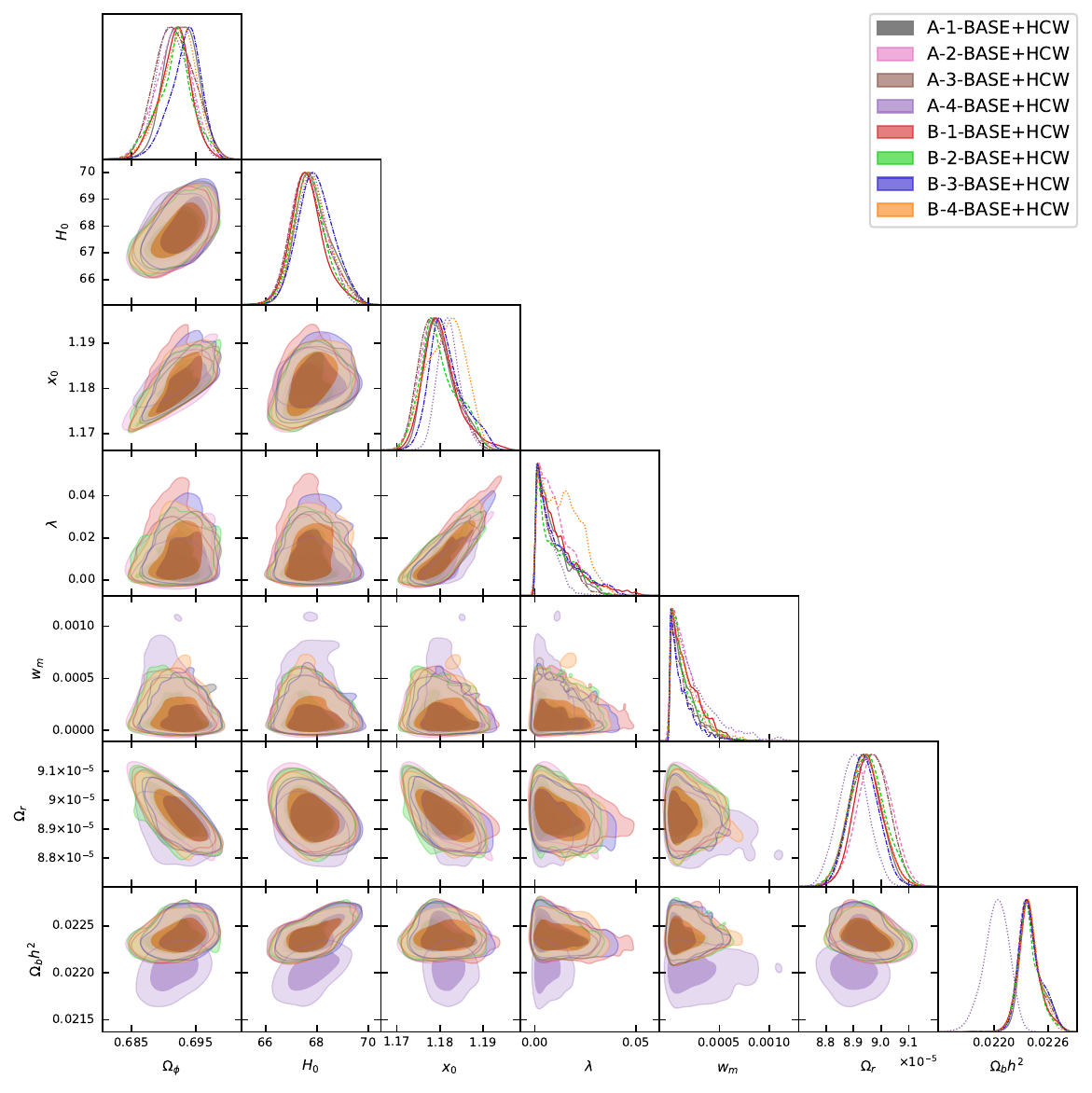}
		\includegraphics[scale=0.4]{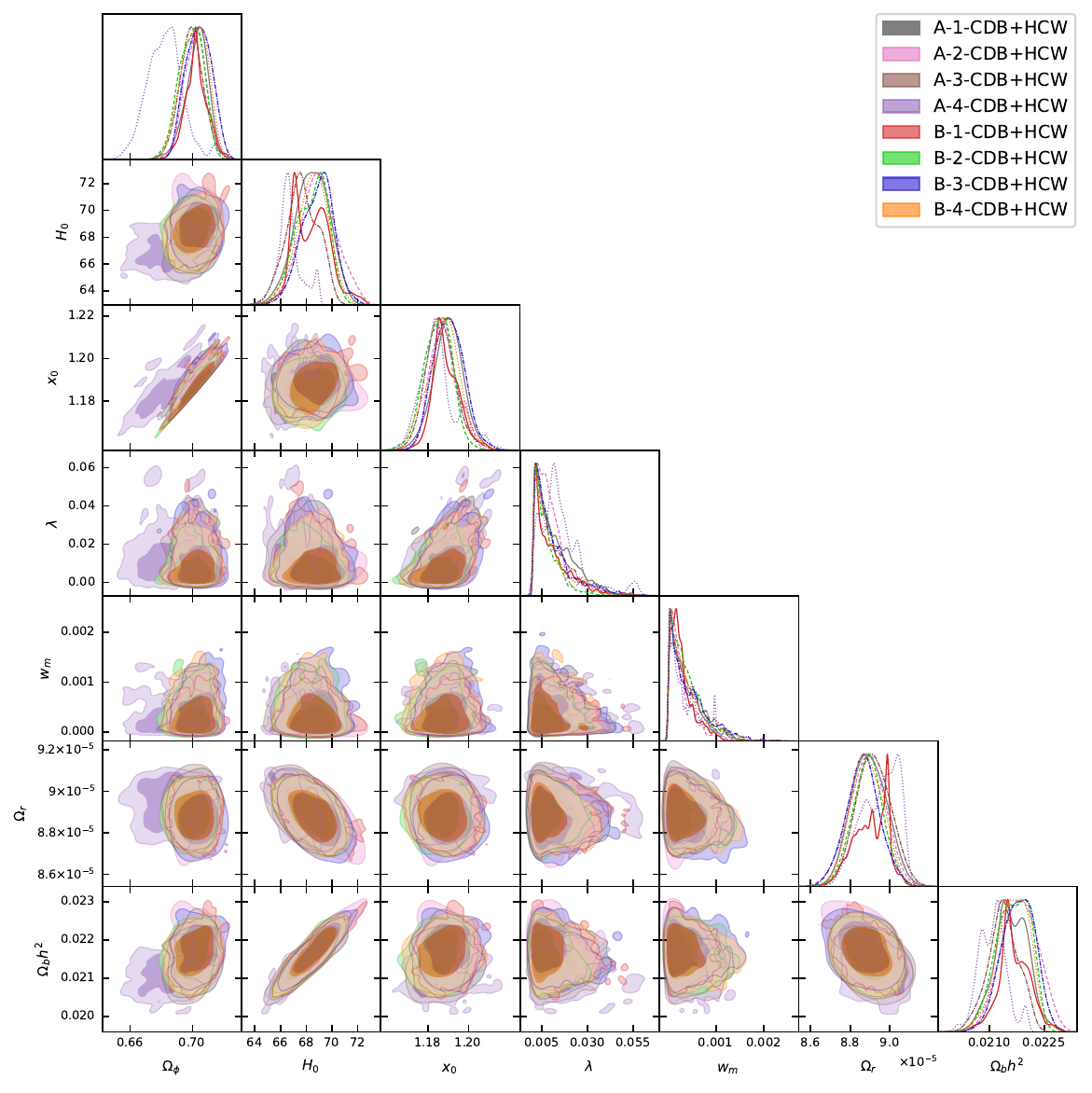}
		
		\includegraphics[scale=0.45]{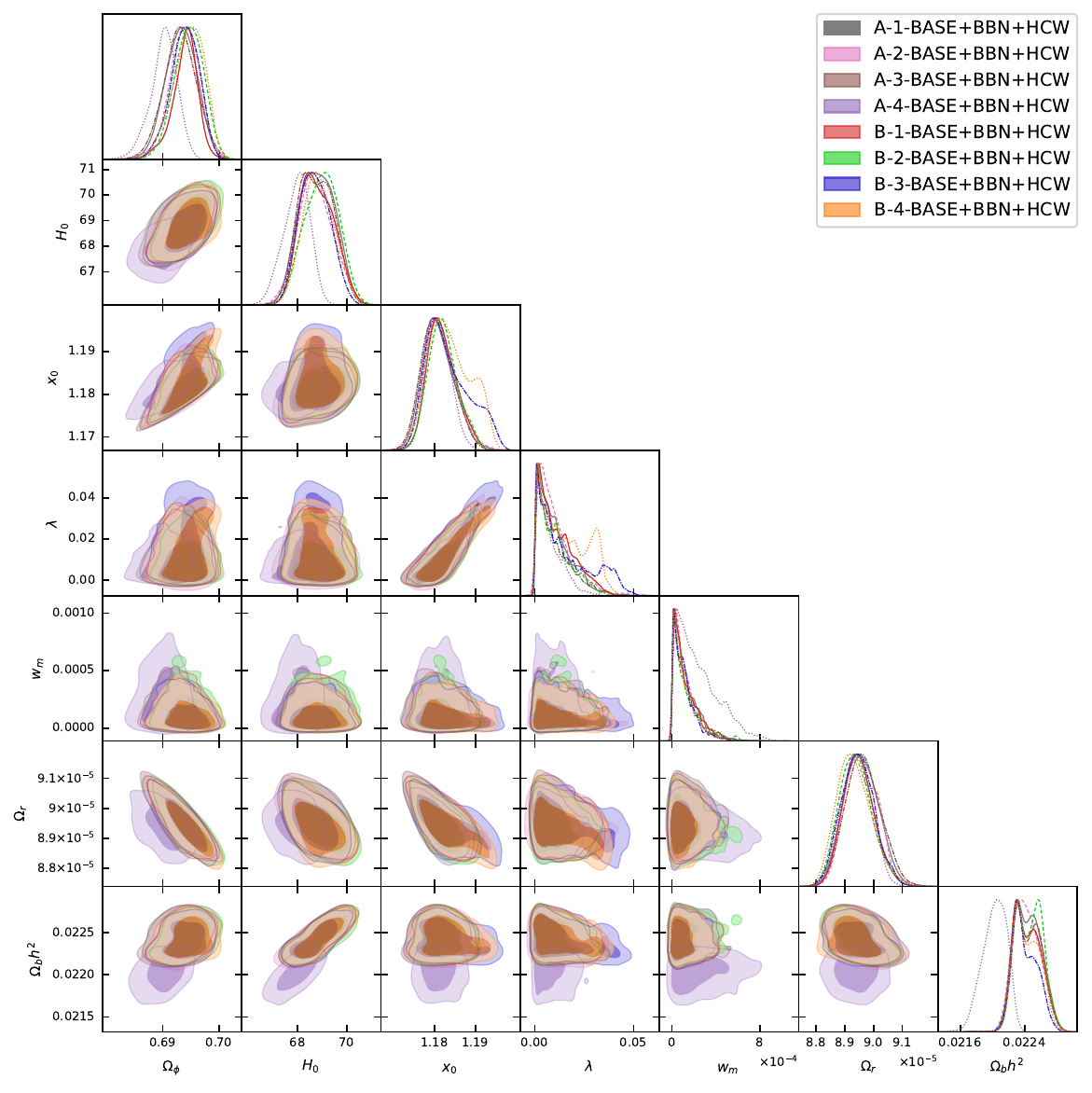}
		
		\caption{Marginalized posterior distributions for the BASE+H0LiCOW, CDB+H0LiCOW, and BASE+BBN+H0LiCOW datasets. Here, BASE and CDB refer to CC+BAO+PLA and CC+BAO+BBN, respectively.}
		
		\label{fig:holicow_chains}
	\end{figure*}

	\section{Conclusion \label{sec:conclusion}}
	
	In this work, we explored classes of interacting models between dark matter (DM) and dark energy (DE). A $k$-essence scalar field with an inverse-square type potential and quadratic kinetic function \(F(X)\) was considered to act as dark energy. A non-zero equation of state \(w\) was assigned to the dark matter fluid to test its deviation from the pressureless nature. Two general classes of interacting models were considered, Model A and Model B, constructed using the time-dependent Hubble parameter, DM-DE energy density and pressure, and the Hubble constant. These interactions were studied by dividing each model into sub-classes where only one term is active at a time. This simplification reduces mathematical complexity, increases numerical stability, and provides a clear framework to understand the behavior of each interaction term. 
	
	To study these interacting models, we set up an autonomous system of equations within the dynamical stability framework. In Model A, since the interaction \(Q\) is proportional to the time-dependent Hubble parameter \(H\), the autonomous system closes using predefined dimensionless variables, yielding a four-dimensional phase space. In Model B, where \(Q \propto H_0\), an additional dynamical variable was introduced to capture the Hubble dynamics, resulting in a five-dimensional phase space. The models were constrained using a Bayesian MCMC analysis with multiple datasets, including cosmic chronometers (CC), DESI BAO (BAO), Compressed Planck (PLA), DES 5Y supernovae (DES), Pantheon+ (PP), Big Bang Nucleosynthesis (BBN), and H0LiCOW (HCW). The baseline dataset was taken as CC+BAO+PLA (BASE), since late-time data alone cannot tightly constrain cosmological parameters. The datasets were combined as BASE+PP (or DES) for supernovae and as BASE+HCW, BASE+BBN+HCW, and CC+BAO+BBN+HCW (CDB+HCW) for strong lensing, chosen according to compatibility and systematics.
	
	Testing the models with supernova samples, nearly all models yielded \(H_0 \sim 66-67\) km/s/Mpc with PP, while DES compilation produced \(H_0 \sim 69.61\) km/s/Mpc, showing a \(\sim 4\sigma\) tension. For strong lensing, the models yielded \(H_0 \sim 67.7\) km/s/Mpc with BASE+HCW, corresponding to \(\sim 2.7\sigma\) tension with DES. CDB+HCW produced \(H_0 \sim 68.5\) km/s/Mpc, reducing the tension, though Model A-IV remained aligned with PP data. Including PLA in BASE+BBN+HCW resulted in a consistent \(H_0 \sim 68.8\) km/s/Mpc, providing an intermediate value between PP and DES.
	
	The analysis shows that the models cannot reach \(H_0 > 71\) km/s/Mpc for the considered datasets, indicating a persistent tension with the SH0ES measurement. Upon evolving the cosmological parameters, all models systematically reproduce the three cosmic phases (radiation, matter, dark energy), showing consistent high-redshift behavior. The scalar field equation of state asymptotically saturates to \(-1\) at late times with {the sound speed approaching very small positive values}, revealing a stable de-Sitter--like universe at the background level. The models yield similar chi-squared values as flat \(\Lambda\)CDM, with some datasets even showing lower values, offering a competitive alternative. {Since the field equation of state approaches \(-1\) without crossing the phantom divide and the sound speed remains positive, the models satisfy a necessary consistency condition related to the absence of ghost-like behavior.}
	
	The interaction function exhibits different behaviors depending on its functional form. For Models A-I to A-III, where the interaction depends on the time-dependent Hubble parameter, the interaction grows monotonically with redshift. When the interaction depends on the field pressure, it shows a transitional behavior around \(N<-3\), preventing rapid growth at high redshift. This transition changes the direction of energy flow between DM and DE, statistically outperforming other models. In Model B, the interaction is independent of the Hubble parameter, so no rapid growth occurs. It shows sudden jumps in intermediate redshift for B-I and B-III, while B-II and B-IV jump at very low redshift. Despite these variations, the energy density evolution remains consistent across all models.
	
	Finally, the interacting models explored here provide consistent results across all analyses. They yield competitive statistical performance, very close to $\Lambda$CDM, and {in some cases slightly better, indicating that a subset of the interacting scenarios considered can serve as a viable alternative at the background level. These results suggest that dynamical dark energy remains a phenomenologically interesting possibility, without implying a definitive preference over a cosmological constant.} Despite this excellent fit, the models considered here, with the assumed datasets, are unable to fully resolve the discrepancy in Hubble measurements. Future work could involve extending these models to include perturbation-level analyses and the full CMB likelihood, which may help reconcile tensions between different observational probes.

	\begin{acknowledgements}
		This work is supported by the National Natural Science Foundation of China under Grants No.~12275238 and No.~W2433018, the National Key Research and Development Program of China under Grant No. 2020YFC2201503, the Zhejiang Provincial Natural Science Foundation of China under Grants No.~LR21A050001 and No.~LY20A050002, and the Fundamental Research Funds for the Provincial Universities of Zhejiang in China under Grant No.~RF-A2019015.
	\end{acknowledgements}

	 \bibliographystyle{JHEP}
	\bibliography{ref} 

@article{SupernovaCosmologyProject:1998vns,
	author = "Perlmutter, S. and others",
	collaboration = "Supernova Cosmology Project",
	title = "{Measurements of $\Omega$ and $\Lambda$ from 42 high redshift supernovae}",
	eprint = "astro-ph/9812133",
	archivePrefix = "arXiv",
	reportNumber = "LBNL-41801, LBL-41801",
	doi = "10.1086/307221",
	journal = "Astrophys. J.",
	volume = "517",
	pages = "565--586",
	year = "1999"
}

@article{SupernovaSearchTeam:1998fmf,
	author = "Riess, Adam G. and others",
	collaboration = "Supernova Search Team",
	title = "{Observational evidence from supernovae for an accelerating universe and a cosmological constant}",
	eprint = "astro-ph/9805201",
	archivePrefix = "arXiv",
	doi = "10.1086/300499",
	journal = "Astron. J.",
	volume = "116",
	pages = "1009--1038",
	year = "1998"
}

@article{WMAP:2003elm,
	author = "Spergel, D. N. and others",
	collaboration = "WMAP",
	title = "{First year Wilkinson Microwave Anisotropy Probe (WMAP) observations: Determination of cosmological parameters}",
	eprint = "astro-ph/0302209",
	archivePrefix = "arXiv",
	doi = "10.1086/377226",
	journal = "Astrophys. J. Suppl.",
	volume = "148",
	pages = "175--194",
	year = "2003"
}

@article{Sherwin:2011gv,
	author = "Sherwin, Blake D. and others",
	title = "{Evidence for dark energy from the cosmic microwave background alone using the Atacama Cosmology Telescope lensing measurements}",
	eprint = "1105.0419",
	archivePrefix = "arXiv",
	primaryClass = "astro-ph.CO",
	doi = "10.1103/PhysRevLett.107.021302",
	journal = "Phys. Rev. Lett.",
	volume = "107",
	pages = "021302",
	year = "2011"
}

@article{Wright:2007vr,
	author = "Wright, Edward L.",
	title = "{Constraints on Dark Energy from Supernovae, Gamma Ray Bursts, Acoustic Oscillations, Nucleosynthesis and Large Scale Structure and the Hubble constant}",
	eprint = "astro-ph/0701584",
	archivePrefix = "arXiv",
	doi = "10.1086/519274",
	journal = "Astrophys. J.",
	volume = "664",
	pages = "633--639",
	year = "2007"
}

@article{DES:2016qvw,
	author = "Kwan, J. and others",
	collaboration = "DES",
	title = "{Cosmology from large-scale galaxy clustering and galaxy\textendash{}galaxy lensing with Dark Energy Survey Science Verification data}",
	eprint = "1604.07871",
	archivePrefix = "arXiv",
	primaryClass = "astro-ph.CO",
	reportNumber = "FERMILAB-PUB-16-146-A-AE",
	doi = "10.1093/mnras/stw2464",
	journal = "Mon. Not. Roy. Astron. Soc.",
	volume = "464",
	number = "4",
	pages = "4045--4062",
	year = "2017"
}

@article{DES:2021esc,
	author = "Abbott, T. M. C. and others",
	collaboration = "DES",
	title = "{Dark Energy Survey Year 3 results: A 2.7\% measurement of baryon acoustic oscillation distance scale at redshift 0.835}",
	eprint = "2107.04646",
	archivePrefix = "arXiv",
	primaryClass = "astro-ph.CO",
	reportNumber = "FERMILAB-PUB-21-838-PPD, DES-2021-0651",
	doi = "10.1103/PhysRevD.105.043512",
	journal = "Phys. Rev. D",
	volume = "105",
	number = "4",
	pages = "043512",
	year = "2022"
}

@article{SDSS:2005xqv,
	author = "Eisenstein, Daniel J. and others",
	collaboration = "SDSS",
	title = "{Detection of the Baryon Acoustic Peak in the Large-Scale Correlation Function of SDSS Luminous Red Galaxies}",
	eprint = "astro-ph/0501171",
	archivePrefix = "arXiv",
	reportNumber = "FERMILAB-PUB-05-057-A-CD",
	doi = "10.1086/466512",
	journal = "Astrophys. J.",
	volume = "633",
	pages = "560--574",
	year = "2005"
}

@article{ACTPol:2014pbf,
	author = "Naess, Sigurd and others",
	collaboration = "ACTPol",
	title = "{The Atacama Cosmology Telescope: CMB Polarization at $200<\ell<9000$}",
	eprint = "1405.5524",
	archivePrefix = "arXiv",
	primaryClass = "astro-ph.CO",
	doi = "10.1088/1475-7516/2014/10/007",
	journal = "JCAP",
	volume = "10",
	pages = "007",
	year = "2014"
}

@article{Planck:2015fie,
	author = "Ade, P. A. R. and others",
	collaboration = "Planck",
	title = "{Planck 2015 results. XIII. Cosmological parameters}",
	eprint = "1502.01589",
	archivePrefix = "arXiv",
	primaryClass = "astro-ph.CO",
	doi = "10.1051/0004-6361/201525830",
	journal = "Astron. Astrophys.",
	volume = "594",
	pages = "A13",
	year = "2016"
}

@article{SDSS:1999zww,
	author = "Fischer, Philippe and others",
	collaboration = "SDSS",
	title = "{Weak lensing with SDSS commissioning data: The Galaxy mass correlation function to $1h^{-1}$ Mpc}",
	eprint = "astro-ph/9912119",
	archivePrefix = "arXiv",
	reportNumber = "FERMILAB-PUB-00-264-A",
	doi = "10.1086/301540",
	journal = "Astron. J.",
	volume = "120",
	pages = "1198--1208",
	year = "2000"
}

@article{Sandage:2006cv,
	author = "Sandage, A. and Tammann, G. A. and Saha, A. and Reindl, B. and Macchetto, F. D. and Panagia, N.",
	title = "{The Hubble Constant: A Summary of the HST Program for the Luminosity Calibration of Type Ia Supernovae by Means of Cepheids}",
	eprint = "astro-ph/0603647",
	archivePrefix = "arXiv",
	doi = "10.1086/508853",
	journal = "Astrophys. J.",
	volume = "653",
	pages = "843--860",
	year = "2006"
}

@inproceedings{Primack:1997av,
	author = "Primack, Joel R.",
	title = "{Dark matter and structure formation}",
	booktitle = "{Midrasha Mathematicae in Jerusalem: Winter School in Dynamical Systems}",
	eprint = "astro-ph/9707285",
	archivePrefix = "arXiv",
	reportNumber = "SCIPP-96-59-REV, SCIPP-96-59",
	month = "7",
	year = "1997"
}

@article{DelPopolo:2007dna,
	author = "Del Popolo, Antonino",
	title = "{Dark matter and structure formation a review}",
	eprint = "0801.1091",
	archivePrefix = "arXiv",
	primaryClass = "astro-ph",
	doi = "10.1134/S1063772907030018",
	journal = "Astron. Rep.",
	volume = "51",
	pages = "169--196",
	year = "2007"
}

@article{Diao:2023tor,
	author = "Diao, Junwen and Wei, Shibiao and Wei, Zherui and Liu, Chang",
	title = "{The impact of the dark matter on galaxy formation}",
	doi = "10.1088/1742-6596/2441/1/012025",
	journal = "J. Phys. Conf. Ser.",
	volume = "2441",
	number = "1",
	pages = "012025",
	year = "2023"
}

@article{Mina:2020eik,
	author = "Mina, Mattia and Mota, David F. and Winther, Hans A.",
	title = "{Solitons in the dark: First approach to non-linear structure formation with fuzzy dark matter}",
	eprint = "2007.04119",
	archivePrefix = "arXiv",
	primaryClass = "astro-ph.CO",
	doi = "10.1051/0004-6361/202038876",
	journal = "Astron. Astrophys.",
	volume = "662",
	pages = "A29",
	year = "2022"
}

@article{Blumenthal:1984bp,
	author = "Blumenthal, George R. and Faber, S. M. and Primack, Joel R. and Rees, Martin J.",
	editor = "Srednicki, M. A.",
	title = "{Formation of Galaxies and Large Scale Structure with Cold Dark Matter}",
	reportNumber = "SLAC-PUB-3307",
	doi = "10.1038/311517a0",
	journal = "Nature",
	volume = "311",
	pages = "517--525",
	year = "1984"
}

@article{Copeland:2006wr,
	author = "Copeland, Edmund J. and Sami, M. and Tsujikawa, Shinji",
	title = "{Dynamics of dark energy}",
	eprint = "hep-th/0603057",
	archivePrefix = "arXiv",
	doi = "10.1142/S021827180600942X",
	journal = "Int. J. Mod. Phys. D",
	volume = "15",
	pages = "1753--1936",
	year = "2006"
}

@article{Weinberg:1988cp,
	author = "Weinberg, Steven",
	editor = "Hsu, Jong-Ping and Fine, D.",
	title = "{The Cosmological Constant Problem}",
	reportNumber = "UTTG-12-88",
	doi = "10.1103/RevModPhys.61.1",
	journal = "Rev. Mod. Phys.",
	volume = "61",
	pages = "1--23",
	year = "1989"
}

@article{Rugh:2000ji,
	author = "Rugh, S. E. and Zinkernagel, H.",
	title = "{The Quantum vacuum and the cosmological constant problem}",
	eprint = "hep-th/0012253",
	archivePrefix = "arXiv",
	doi = "10.1016/S1355-2198(02)00033-3",
	journal = "Stud. Hist. Phil. Sci. B",
	volume = "33",
	pages = "663--705",
	year = "2002"
}

@article{Padmanabhan:2002ji,
	author = "Padmanabhan, T.",
	title = "{Cosmological constant: The Weight of the vacuum}",
	eprint = "hep-th/0212290",
	archivePrefix = "arXiv",
	doi = "10.1016/S0370-1573(03)00120-0",
	journal = "Phys. Rept.",
	volume = "380",
	pages = "235--320",
	year = "2003"
}

@article{Carroll:1991mt,
	author = "Carroll, Sean M. and Press, William H. and Turner, Edwin L.",
	title = "{The Cosmological constant}",
	reportNumber = "CFA-3332",
	doi = "10.1146/annurev.aa.30.090192.002435",
	journal = "Ann. Rev. Astron. Astrophys.",
	volume = "30",
	pages = "499--542",
	year = "1992"
}

@article{Planck:2018vyg,
	author = "Aghanim, N. and others",
	collaboration = "Planck",
	title = "{Planck 2018 results. VI. Cosmological parameters}",
	eprint = "1807.06209",
	archivePrefix = "arXiv",
	primaryClass = "astro-ph.CO",
	doi = "10.1051/0004-6361/201833910",
	journal = "Astron. Astrophys.",
	volume = "641",
	pages = "A6",
	year = "2020",
	note = "[Erratum: Astron.Astrophys. 652, C4 (2021)]"
}

@article{Riess:2020fzl,
	author = "Riess, Adam G. and Casertano, Stefano and Yuan, Wenlong and Bowers, J. Bradley and Macri, Lucas and Zinn, Joel C. and Scolnic, Dan",
	title = "{Cosmic Distances Calibrated to 1\% Precision with Gaia EDR3 Parallaxes and Hubble Space Telescope Photometry of 75 Milky Way Cepheids Confirm Tension with $\Lambda$CDM}",
	eprint = "2012.08534",
	archivePrefix = "arXiv",
	primaryClass = "astro-ph.CO",
	doi = "10.3847/2041-8213/abdbaf",
	journal = "Astrophys. J. Lett.",
	volume = "908",
	number = "1",
	pages = "L6",
	year = "2021"
}

@article{Freedman:2020dne,
	author = "Freedman, Wendy L. and Madore, Barry F. and Hoyt, Taylor and Jang, In Sung and Beaton, Rachael and Lee, Myung Gyoon and Monson, Andrew and Neeley, Jill and Rich, Jeffrey",
	title = "{Calibration of the Tip of the Red Giant Branch (TRGB)}",
	eprint = "2002.01550",
	archivePrefix = "arXiv",
	primaryClass = "astro-ph.GA",
	doi = "10.3847/1538-4357/ab7339",
	month = "2",
	year = "2020",
	journal=""
}

@article{Pesce:2020xfe,
	author = "Pesce, D. W. and others",
	title = "{The Megamaser Cosmology Project. XIII. Combined Hubble constant constraints}",
	eprint = "2001.09213",
	archivePrefix = "arXiv",
	primaryClass = "astro-ph.CO",
	doi = "10.3847/2041-8213/ab75f0",
	journal = "Astrophys. J. Lett.",
	volume = "891",
	number = "1",
	pages = "L1",
	year = "2020"
}

@article{SPT-3G:2021eoc,
	author = "Dutcher, D. and others",
	collaboration = "SPT-3G",
	title = "{Measurements of the E-mode polarization and temperature-E-mode correlation of the CMB from SPT-3G 2018 data}",
	eprint = "2101.01684",
	archivePrefix = "arXiv",
	primaryClass = "astro-ph.CO",
	reportNumber = "FERMILAB-PUB-21-137-AE",
	doi = "10.1103/PhysRevD.104.022003",
	journal = "Phys. Rev. D",
	volume = "104",
	number = "2",
	pages = "022003",
	year = "2021"
}

@article{Cuceu:2019for,
	author = "Cuceu, Andrei and Farr, James and Lemos, Pablo and Font-Ribera, Andreu",
	title = "{Baryon Acoustic Oscillations and the Hubble Constant: Past, Present and Future}",
	eprint = "1906.11628",
	archivePrefix = "arXiv",
	primaryClass = "astro-ph.CO",
	doi = "10.1088/1475-7516/2019/10/044",
	journal = "JCAP",
	volume = "10",
	pages = "044",
	year = "2019"
}

@article{Pascale:2024qjr,
	author = "Pascale, Massimo and others",
	title = "{SN H0pe: The First Measurement of H$_{0}$ from a Multiply Imaged Type Ia Supernova, Discovered by JWST}",
	eprint = "2403.18902",
	archivePrefix = "arXiv",
	primaryClass = "astro-ph.CO",
	doi = "10.3847/1538-4357/ad9928",
	journal = "Astrophys. J.",
	volume = "979",
	number = "1",
	pages = "13",
	year = "2025"
}

@article{DESI:2024mwx,
	author = "Adame, A. G. and others",
	collaboration = "DESI",
	title = "{DESI 2024 VI: cosmological constraints from the measurements of baryon acoustic oscillations}",
	eprint = "2404.03002",
	archivePrefix = "arXiv",
	primaryClass = "astro-ph.CO",
	reportNumber = "FERMILAB-PUB-24-0154-PPD",
	doi = "10.1088/1475-7516/2025/02/021",
	journal = "JCAP",
	volume = "02",
	pages = "021",
	year = "2025"
}

@article{Cortes:2024lgw,
	author = "Cort{\^e}s, Marina and Liddle, Andrew R.",
	title = "{Interpreting DESI's evidence for evolving dark energy}",
	eprint = "2404.08056",
	archivePrefix = "arXiv",
	primaryClass = "astro-ph.CO",
	doi = "10.1088/1475-7516/2024/12/007",
	journal = "JCAP",
	volume = "12",
	pages = "007",
	year = "2024"
}

@article{DESI:2025fii,
	author = "Lodha, K. and others",
	collaboration = "DESI",
	title = "{Extended Dark Energy analysis using DESI DR2 BAO measurements}",
	eprint = "2503.14743",
	archivePrefix = "arXiv",
	primaryClass = "astro-ph.CO",
	reportNumber = "FERMILAB-PUB-25-0164-PPD",
	month = "3",
	year = "2025",
	doi = " "
}

@article{Hussain:2025nqy,
	author = "Hussain, Saddam and Arora, Simran and Wang, Anzhong and Rose, Ben",
	title = "{Probing the Dynamics of Gaussian Dark Energy Equation of State Using DESI BAO}",
	eprint = "2505.09913",
	archivePrefix = "arXiv",
	primaryClass = "astro-ph.CO",
	month = "5",
	year = "2025",
	doi=" "
}

@article{Arora:2025msq,
	author = "Arora, Simran and De Felice, Antonio and Mukohyama, Shinji",
	title = "{Dynamical dark energy parameterizations in VCDM}",
	eprint = "2508.03784",
	archivePrefix = "arXiv",
	primaryClass = "gr-qc",
	month = "8",
	year = "2025",
	doi =" "
}

@article{Scherer:2025esj,
	author = "Scherer, Mateus and Sabogal, Miguel A. and Nunes, Rafael C. and De Felice, Antonio",
	title = "{Challenging the {\ensuremath{\Lambda}}CDM model: 5{\ensuremath{\sigma}} evidence for a dynamical dark energy late-time transition}",
	eprint = "2504.20664",
	archivePrefix = "arXiv",
	primaryClass = "astro-ph.CO",
	doi = "10.1103/n86r-sjgm",
	journal = "Phys. Rev. D",
	volume = "112",
	number = "4",
	pages = "043513",
	year = "2025"
}

@article{DESI:2025zgx,
	author = "Abdul Karim, M. and others",
	collaboration = "DESI",
	title = "{DESI DR2 Results II: Measurements of Baryon Acoustic Oscillations and Cosmological Constraints}",
	eprint = "2503.14738",
	archivePrefix = "arXiv",
	primaryClass = "astro-ph.CO",
	reportNumber = "FERMILAB-PUB-25-0169-PPD",
	month = "3",
	year = "2025",
	doi =" "
}

@article{Lima:2004cq,
	author = "Lima, Jose Ademir Sales",
	editor = "Gay Ducati, M. B.",
	title = "{Alternative dark energy models: An Overview}",
	eprint = "astro-ph/0402109",
	archivePrefix = "arXiv",
	doi = "10.1590/S0103-97332004000200009",
	journal = "Braz. J. Phys.",
	volume = "34",
	number = "1A",
	pages = "194--200",
	year = "2004"
}

@article{Wu:2025wyk,
	author = "Wu, Peng-Ju",
	title = "{Comparison of dark energy models using late-universe observations}",
	eprint = "2504.09054",
	archivePrefix = "arXiv",
	primaryClass = "astro-ph.CO",
	month = "4",
	year = "2025",
	doi= " "
}

@article{Clifton:2011jh,
	author = "Clifton, Timothy and Ferreira, Pedro G. and Padilla, Antonio and Skordis, Constantinos",
	title = "{Modified Gravity and Cosmology}",
	eprint = "1106.2476",
	archivePrefix = "arXiv",
	primaryClass = "astro-ph.CO",
	doi = "10.1016/j.physrep.2012.01.001",
	journal = "Phys. Rept.",
	volume = "513",
	pages = "1--189",
	year = "2012"
}

@article{Gomez-Valent:2023hov,
	author = "G{\'o}mez-Valent, Adri{\`a} and Mavromatos, Nick E. and Sol{\`a} Peracaula, Joan",
	title = "{Stringy running vacuum model and current tensions in cosmology}",
	eprint = "2305.15774",
	archivePrefix = "arXiv",
	primaryClass = "gr-qc",
	doi = "10.1088/1361-6382/ad0fb8",
	journal = "Class. Quant. Grav.",
	volume = "41",
	number = "1",
	pages = "015026",
	year = "2024"
}

@article{Giani:2024nnv,
	author = "Giani, Leonardo and Von Marttens, Rodrigo and Camilleri, Ryan",
	title = "{Novel Approach to Cosmological Nonlinearities as an Effective Fluid}",
	eprint = "2410.15295",
	archivePrefix = "arXiv",
	primaryClass = "astro-ph.CO",
	doi = "10.1103/zr92-m7py",
	journal = "Phys. Rev. Lett.",
	volume = "135",
	number = "7",
	pages = "071004",
	year = "2025"
}

@article{Zimdahl:2001ar,
	author = "Zimdahl, Winfried and Pavon, Diego",
	title = "{Interacting quintessence}",
	eprint = "astro-ph/0105479",
	archivePrefix = "arXiv",
	doi = "10.1016/S0370-2693(01)01174-1",
	journal = "Phys. Lett. B",
	volume = "521",
	pages = "133--138",
	year = "2001"
}

@article{Bertacca:2010ct,
	author = "Bertacca, Daniele and Bartolo, Nicola and Matarrese, Sabino",
	title = "{Unified Dark Matter Scalar Field Models}",
	eprint = "1008.0614",
	archivePrefix = "arXiv",
	primaryClass = "astro-ph.CO",
	doi = "10.1155/2010/904379",
	journal = "Adv. Astron.",
	volume = "2010",
	pages = "904379",
	year = "2010"
}

@article{Ansoldi:2012pi,
	author = "Ansoldi, Stefano and Guendelman, Eduardo I.",
	title = "{Unified Dark Energy-Dark Matter model with Inverse Quintessence}",
	eprint = "1209.4758",
	archivePrefix = "arXiv",
	primaryClass = "gr-qc",
	doi = "10.1088/1475-7516/2013/05/036",
	journal = "JCAP",
	volume = "05",
	pages = "036",
	year = "2013"
}

@article{Yao:2024kex,
	author = "Yao, Yan-Hong and Liu, Jian-Qi and Huang, Zhi-Qi and Wang, Jun-Chao and Su, Yan",
	title = "{New unified dark sector model and its implications on the {\ensuremath{\sigma}}8 and S8 tensions}",
	eprint = "2409.04678",
	archivePrefix = "arXiv",
	primaryClass = "astro-ph.CO",
	doi = "10.1103/p51k-7rw2",
	journal = "Phys. Rev. D",
	volume = "111",
	number = "12",
	pages = "123508",
	year = "2025"
}

@article{Gross:1986mw,
	author = "Gross, David J. and Sloan, John H.",
	title = "{The Quartic Effective Action for the Heterotic String}",
	reportNumber = "NSF-ITP-87-02",
	doi = "10.1016/0550-3213(87)90465-2",
	journal = "Nucl. Phys. B",
	volume = "291",
	pages = "41--89",
	year = "1987"
}

@article{Bento:1995qc,
	author = "Bento, M. C. and Bertolami, O.",
	title = "{Maximally Symmetric Cosmological Solutions of higher curvature string effective theories with dilatons}",
	eprint = "gr-qc/9503057",
	archivePrefix = "arXiv",
	reportNumber = "CERN-TH-95-63, CERN-TH-95-063, DFTT-19-95",
	doi = "10.1016/0370-2693(95)01519-1",
	journal = "Phys. Lett. B",
	volume = "368",
	pages = "198--201",
	year = "1996"
}

@article{Nojiri:2005vv,
	author = "Nojiri, Shin'ichi and Odintsov, Sergei D. and Sasaki, Misao",
	title = "{Gauss-Bonnet dark energy}",
	eprint = "hep-th/0504052",
	archivePrefix = "arXiv",
	reportNumber = "YITP-05-14",
	doi = "10.1103/PhysRevD.71.123509",
	journal = "Phys. Rev. D",
	volume = "71",
	pages = "123509",
	year = "2005"
}

@article{Tsujikawa:2006ph,
	author = "Tsujikawa, Shinji and Sami, M.",
	title = "{String-inspired cosmology: Late time transition from scaling matter era to dark energy universe caused by a Gauss-Bonnet coupling}",
	eprint = "hep-th/0608178",
	archivePrefix = "arXiv",
	doi = "10.1088/1475-7516/2007/01/006",
	journal = "JCAP",
	volume = "01",
	pages = "006",
	year = "2007"
}

@article{Hussain:2024yee,
	author = "Hussain, Saddam and Arora, Simran and Rana, Yamuna and Rose, Benjamin and Wang, Anzhong",
	title = "{Interacting models of dark energy and dark matter in Einstein scalar Gauss Bonnet gravity}",
	eprint = "2408.05484",
	archivePrefix = "arXiv",
	primaryClass = "gr-qc",
	doi = "10.1088/1475-7516/2024/11/042",
	journal = "JCAP",
	volume = "11",
	pages = "042",
	year = "2024"
}

@article{Hussain:2025vbo,
	author = "Hussain, Saddam and Arora, Simran and Rana, Yamuna and Rose, Benjamin and Wang, Anzhong",
	title = "{Interacting Scalar Fields as Dark Energy and Dark Matter in Einstein scalar Gauss Bonnet Gravity}",
	eprint = "2507.05207",
	archivePrefix = "arXiv",
	primaryClass = "gr-qc",
	month = "7",
	year = "2025",
	dot= " "
}

@article{Odintsov:2019clh,
	author = "Odintsov, S. D. and Oikonomou, V. K.",
	title = "{Inflationary Phenomenology of Einstein Gauss-Bonnet Gravity Compatible with GW170817}",
	eprint = "1908.07555",
	archivePrefix = "arXiv",
	primaryClass = "gr-qc",
	doi = "10.1016/j.physletb.2019.134874",
	journal = "Phys. Lett. B",
	volume = "797",
	pages = "134874",
	year = "2019"
}

@article{Peebles:2002gy,
	author = "Peebles, P. J. E. and Ratra, Bharat",
	editor = "Hsu, Jong-Ping and Fine, D.",
	title = "{The Cosmological Constant and Dark Energy}",
	eprint = "astro-ph/0207347",
	archivePrefix = "arXiv",
	reportNumber = "KSUPT-02-3",
	doi = "10.1103/RevModPhys.75.559",
	journal = "Rev. Mod. Phys.",
	volume = "75",
	pages = "559--606",
	year = "2003"
}

@article{Nishioka:1992sg,
	author = "Nishioka, T. and Fujii, Y.",
	title = "{Inflation and the decaying cosmological constant}",
	doi = "10.1103/PhysRevD.45.2140",
	journal = "Phys. Rev. D",
	volume = "45",
	pages = "2140--2143",
	year = "1992"
}

@article{Ferreira:1997hj,
	author = "Ferreira, Pedro G. and Joyce, Michael",
	title = "{Cosmology with a primordial scaling field}",
	eprint = "astro-ph/9711102",
	archivePrefix = "arXiv",
	reportNumber = "CFPA-97-TH-20",
	doi = "10.1103/PhysRevD.58.023503",
	journal = "Phys. Rev. D",
	volume = "58",
	pages = "023503",
	year = "1998"
}

@article{Copeland:1997et,
	author = "Copeland, Edmund J. and Liddle, Andrew R and Wands, David",
	title = "{Exponential potentials and cosmological scaling solutions}",
	eprint = "gr-qc/9711068",
	archivePrefix = "arXiv",
	reportNumber = "SUSX-TH-97-022, SUSSEX-AST-97-11-1, PU-RCG-97-20",
	doi = "10.1103/PhysRevD.57.4686",
	journal = "Phys. Rev. D",
	volume = "57",
	pages = "4686--4690",
	year = "1998"
}

@article{Armendariz-Picon:2000nqq,
	author = "Armendariz-Picon, C. and Mukhanov, Viatcheslav F. and Steinhardt, Paul J.",
	title = "{A Dynamical solution to the problem of a small cosmological constant and late time cosmic acceleration}",
	eprint = "astro-ph/0004134",
	archivePrefix = "arXiv",
	doi = "10.1103/PhysRevLett.85.4438",
	journal = "Phys. Rev. Lett.",
	volume = "85",
	pages = "4438--4441",
	year = "2000"
}

@article{Armendariz-Picon:2000ulo,
	author = "Armendariz-Picon, C. and Mukhanov, Viatcheslav F. and Steinhardt, Paul J.",
	title = "{Essentials of k essence}",
	eprint = "astro-ph/0006373",
	archivePrefix = "arXiv",
	doi = "10.1103/PhysRevD.63.103510",
	journal = "Phys. Rev. D",
	volume = "63",
	pages = "103510",
	year = "2001"
}

@article{Chiba:1999ka,
	author = "Chiba, Takeshi and Okabe, Takahiro and Yamaguchi, Masahide",
	title = "{Kinetically driven quintessence}",
	eprint = "astro-ph/9912463",
	archivePrefix = "arXiv",
	reportNumber = "UTAP-352",
	doi = "10.1103/PhysRevD.62.023511",
	journal = "Phys. Rev. D",
	volume = "62",
	pages = "023511",
	year = "2000"
}

@article{Armendariz-Picon:1999hyi,
	author = "Armendariz-Picon, C. and Damour, T. and Mukhanov, Viatcheslav F.",
	title = "{k - inflation}",
	eprint = "hep-th/9904075",
	archivePrefix = "arXiv",
	doi = "10.1016/S0370-2693(99)00603-6",
	journal = "Phys. Lett. B",
	volume = "458",
	pages = "209--218",
	year = "1999"
}

@article{Armendariz-Picon:2005oog,
	author = "Armendariz-Picon, C. and Lim, Eugene A.",
	title = "{Haloes of k-essence}",
	eprint = "astro-ph/0505207",
	archivePrefix = "arXiv",
	doi = "10.1088/1475-7516/2005/08/007",
	journal = "JCAP",
	volume = "08",
	pages = "007",
	year = "2005"
}

@article{Arkani-Hamed:2003pdi,
	author = "Arkani-Hamed, Nima and Cheng, Hsin-Chia and Luty, Markus A. and Mukohyama, Shinji",
	title = "{Ghost condensation and a consistent infrared modification of gravity}",
	eprint = "hep-th/0312099",
	archivePrefix = "arXiv",
	reportNumber = "HUTP-03-A081, UMD-PPP-04-012",
	doi = "10.1088/1126-6708/2004/05/074",
	journal = "JHEP",
	volume = "05",
	pages = "074",
	year = "2004"
}

@article{Scherrer:2004au,
	author = "Scherrer, Robert J.",
	title = "{Purely kinetic k-essence as unified dark matter}",
	eprint = "astro-ph/0402316",
	archivePrefix = "arXiv",
	doi = "10.1103/PhysRevLett.93.011301",
	journal = "Phys. Rev. Lett.",
	volume = "93",
	pages = "011301",
	year = "2004"
}

@article{Chatterjee:2021ijw,
	author = "Chatterjee, Anirban and Hussain, Saddam and Bhattacharya, Kaushik",
	title = "{Dynamical stability of the k-essence field interacting nonminimally with a perfect fluid}",
	eprint = "2105.00361",
	archivePrefix = "arXiv",
	primaryClass = "gr-qc",
	doi = "10.1103/PhysRevD.104.103505",
	journal = "Phys. Rev. D",
	volume = "104",
	number = "10",
	pages = "103505",
	year = "2021"
}

@article{Hussain:2022osn,
	author = "Hussain, Saddam and Chatterjee, Anirban and Bhattacharya, Kaushik",
	title = "{Ghost Condensates and Pure Kinetic k-Essence Condensates in the Presence of Field\textendash{}Fluid Non-Minimal Coupling in the Dark Sector}",
	eprint = "2203.10607",
	archivePrefix = "arXiv",
	primaryClass = "gr-qc",
	doi = "10.3390/universe9020065",
	journal = "Universe",
	volume = "9",
	number = "2",
	pages = "65",
	year = "2023"
}

@article{Bhattacharya:2022wzu,
	author = "Bhattacharya, Kaushik and Chatterjee, Anirban and Hussain, Saddam",
	title = "{Dynamical stability in presence of non-minimal derivative dependent coupling of k-essence field with a relativistic fluid}",
	eprint = "2206.12398",
	archivePrefix = "arXiv",
	primaryClass = "gr-qc",
	doi = "10.1140/epjc/s10052-023-11666-w",
	journal = "Eur. Phys. J. C",
	volume = "83",
	number = "6",
	pages = "488",
	year = "2023"
}

@article{Hussain:2022dhp,
	author = "Hussain, Saddam and Chakraborty, Saikat and Roy, Nandan and Bhattacharya, Kaushik",
	title = "{Dynamical systems analysis of tachyon-dark-energy models from a new perspective}",
	eprint = "2208.10352",
	archivePrefix = "arXiv",
	primaryClass = "gr-qc",
	doi = "10.1103/PhysRevD.107.063515",
	journal = "Phys. Rev. D",
	volume = "107",
	number = "6",
	pages = "063515",
	year = "2023"
}

@article{Fang:2014qga,
	author = "Fang, Wei and Tu, Hong and Li, Ying and Huang, Jiasheng and Shu, Chenggang",
	title = "{Full Investigation on the Dynamics of Power-Law Kinetic Quintessence}",
	eprint = "1406.0128",
	archivePrefix = "arXiv",
	primaryClass = "gr-qc",
	doi = "10.1103/PhysRevD.89.123514",
	journal = "Phys. Rev. D",
	volume = "89",
	number = "12",
	pages = "123514",
	year = "2014"
}

@article{Yang:2009zzl,
	author = "Yang, Rong-Jia and Gao, Xiang-Ting",
	title = "{Observational constraints on purely kinetic k-essence dark energy models}",
	doi = "10.1088/0256-307X/26/8/089501",
	journal = "Chin. Phys. Lett.",
	volume = "26",
	pages = "089501",
	year = "2009"
}

@article{Dinda:2023mad,
	author = "Dinda, Bikash R. and Banerjee, Narayan",
	title = "{Constraints on the speed of sound in the k-essence model of dark energy}",
	eprint = "2309.10538",
	archivePrefix = "arXiv",
	primaryClass = "astro-ph.CO",
	doi = "10.1140/epjc/s10052-024-12547-6",
	journal = "Eur. Phys. J. C",
	volume = "84",
	number = "2",
	pages = "177",
	year = "2024"
}

@article{Hussain:2024qrd,
	author = "Hussain, Saddam and Nelleri, Sarath and Bhattacharya, Kaushik",
	title = "{Comprehensive study of k-essence model: dynamical system analysis and observational constraints from latest Type Ia supernova and BAO observations}",
	eprint = "2406.07179",
	archivePrefix = "arXiv",
	primaryClass = "astro-ph.CO",
	doi = "10.1088/1475-7516/2025/03/025",
	journal = "JCAP",
	volume = "03",
	pages = "025",
	year = "2025"
}

@article{Caldwell:2003vq,
	author = "Caldwell, Robert R. and Kamionkowski, Marc and Weinberg, Nevin N.",
	title = "{Phantom energy and cosmic doomsday}",
	eprint = "astro-ph/0302506",
	archivePrefix = "arXiv",
	doi = "10.1103/PhysRevLett.91.071301",
	journal = "Phys. Rev. Lett.",
	volume = "91",
	pages = "071301",
	year = "2003"
}

@article{Ludwick:2017tox,
	author = "Ludwick, Kevin J.",
	title = "{The viability of phantom dark energy: A review}",
	eprint = "1708.06981",
	archivePrefix = "arXiv",
	primaryClass = "astro-ph.CO",
	doi = "10.1142/S0217732317300257",
	journal = "Mod. Phys. Lett. A",
	volume = "32",
	number = "28",
	pages = "1730025",
	year = "2017"
}

@article{Wang:2016lxa,
	author = "Wang, B. and Abdalla, E. and Atrio-Barandela, F. and Pavon, D.",
	title = "{Dark Matter and Dark Energy Interactions: Theoretical Challenges, Cosmological Implications and Observational Signatures}",
	eprint = "1603.08299",
	archivePrefix = "arXiv",
	primaryClass = "astro-ph.CO",
	doi = "10.1088/0034-4885/79/9/096901",
	journal = "Rept. Prog. Phys.",
	volume = "79",
	number = "9",
	pages = "096901",
	year = "2016"
}

@article{Lee:2006za,
	author = "Lee, Seokcheon and Liu, Guo-Chin and Ng, Kin-Wang",
	title = "{Constraints on the coupled quintessence from cosmic microwave background anisotropy and matter power spectrum}",
	eprint = "astro-ph/0601333",
	archivePrefix = "arXiv",
	doi = "10.1103/PhysRevD.73.083516",
	journal = "Phys. Rev. D",
	volume = "73",
	pages = "083516",
	year = "2006"
}

@article{Bandyopadhyay:2017igc,
	author = "Bandyopadhyay, Abhijit and Chatterjee, Anirban",
	title = "{Realizing interactions between dark matter and dark energy using $k$-essence cosmology}",
	eprint = "1709.04334",
	archivePrefix = "arXiv",
	primaryClass = "gr-qc",
	doi = "10.1142/S0217732319502195",
	journal = "Mod. Phys. Lett. A",
	volume = "34",
	number = "27",
	pages = "1950219",
	year = "2019"
}

@article{Kritpetch:2024rgi,
	author = "Kritpetch, Chonticha and Roy, Nandan and Banerjee, Narayan",
	title = "{Interacting dark sector: A dynamical system perspective}",
	eprint = "2405.10604",
	archivePrefix = "arXiv",
	primaryClass = "gr-qc",
	doi = "10.1103/PhysRevD.111.103501",
	journal = "Phys. Rev. D",
	volume = "111",
	number = "10",
	pages = "103501",
	year = "2025"
}

@article{Li:2025owk,
	author = "Li, Tian-Nuo and Du, Guo-Hong and Li, Yun-He and Wu, Peng-Ju and Jin, Shang-Jie and Zhang, Jing-Fei and Zhang, Xin",
	title = "{Probing the sign-changeable interaction between dark energy and dark matter with DESI baryon acoustic oscillations and DES supernovae data}",
	eprint = "2501.07361",
	archivePrefix = "arXiv",
	primaryClass = "astro-ph.CO",
	month = "1",
	year = "2025",
	doi = " "
}

@article{Garriga:1999vw,
	author = "Garriga, Jaume and Mukhanov, Viatcheslav F.",
	title = "{Perturbations in k-inflation}",
	eprint = "hep-th/9904176",
	archivePrefix = "arXiv",
	reportNumber = "UAB-FT-466",
	doi = "10.1016/S0370-2693(99)00602-4",
	journal = "Phys. Lett. B",
	volume = "458",
	pages = "219--225",
	year = "1999"
}

@article{Linton:2017ged,
	author = "Linton, Mark S. and Pourtsidou, Alkistis and Crittenden, Robert and Maartens, Roy",
	title = "{Variable sound speed in interacting dark energy models}",
	eprint = "1711.05196",
	archivePrefix = "arXiv",
	primaryClass = "astro-ph.CO",
	doi = "10.1088/1475-7516/2018/04/043",
	journal = "JCAP",
	volume = "04",
	pages = "043",
	year = "2018"
}

@article{Valiviita:2008iv,
	author = "Valiviita, Jussi and Majerotto, Elisabetta and Maartens, Roy",
	title = "{Instability in interacting dark energy and dark matter fluids}",
	eprint = "0804.0232",
	archivePrefix = "arXiv",
	primaryClass = "astro-ph",
	doi = "10.1088/1475-7516/2008/07/020",
	journal = "JCAP",
	volume = "07",
	pages = "020",
	year = "2008"
}

@article{Piattella:2013wpa,
	author = "Piattella, Oliver F. and Fabris, J{\'u}lio C. and Bili{\'c}, Neven",
	title = "{Note on the thermodynamics and the speed of sound of a scalar field}",
	eprint = "1309.4282",
	archivePrefix = "arXiv",
	primaryClass = "gr-qc",
	doi = "10.1088/0264-9381/31/5/055006",
	journal = "Class. Quant. Grav.",
	volume = "31",
	pages = "055006",
	year = "2014"
}

@article{Jorge:2007zz,
	author = "Jorge, Pedro and Mimoso, Jose P. and Wands, David",
	editor = "Apostolopoulos, P. and Bona, Carles and Carot, J. and Mas, Ll. and Sintes, A. M. and Stela, J.",
	title = "{On the dynamics of k-essence models}",
	doi = "10.1088/1742-6596/66/1/012031",
	journal = "J. Phys. Conf. Ser.",
	volume = "66",
	pages = "012031",
	year = "2007"
}

@article{Bahamonde:2017ize,
	author = {Bahamonde, Sebastian and B{\"o}hmer, Christian G. and Carloni, Sante and Copeland, Edmund J. and Fang, Wei and Tamanini, Nicola},
	title = "{Dynamical systems applied to cosmology: dark energy and modified gravity}",
	eprint = "1712.03107",
	archivePrefix = "arXiv",
	primaryClass = "gr-qc",
	doi = "10.1016/j.physrep.2018.09.001",
	journal = "Phys. Rept.",
	volume = "775-777",
	pages = "1--122",
	year = "2018"
}

@article{Bouhmadi-Lopez:2016dzw,
	author = "Bouhmadi-L{\'o}pez, Mariam and Marto, Jo{\~a}o and Morais, Jo{\~a}o and Silva, C{\'e}sar M.",
	title = "{Cosmic infinity: A dynamical system approach}",
	eprint = "1611.03100",
	archivePrefix = "arXiv",
	primaryClass = "gr-qc",
	doi = "10.1088/1475-7516/2017/03/042",
	journal = "JCAP",
	volume = "03",
	pages = "042",
	year = "2017"
}

@article{Alho:2020cdg,
	author = "Alho, Artur and Uggla, Claes and Wainwright, John",
	title = "{Dynamical systems in perturbative scalar field cosmology}",
	eprint = "2006.00800",
	archivePrefix = "arXiv",
	primaryClass = "gr-qc",
	doi = "10.1088/1361-6382/abb73a",
	journal = "Class. Quant. Grav.",
	volume = "37",
	number = "22",
	pages = "225011",
	year = "2020"
}

@article{Das:2019ixt,
	author = "Das, Sudipta and Banerjee, Manisha and Roy, Nandan",
	title = "{Dynamical System Analysis for Steep Potentials}",
	eprint = "1903.02288",
	archivePrefix = "arXiv",
	primaryClass = "gr-qc",
	doi = "10.1088/1475-7516/2019/08/024",
	journal = "JCAP",
	volume = "08",
	pages = "024",
	year = "2019"
}

@article{Moresco:2012jh,
	author = "Moresco, M. and others",
	title = "{Improved constraints on the expansion rate of the Universe up to z{\textasciitilde}1.1 from the spectroscopic evolution of cosmic chronometers}",
	eprint = "1201.3609",
	archivePrefix = "arXiv",
	primaryClass = "astro-ph.CO",
	doi = "10.1088/1475-7516/2012/08/006",
	journal = "JCAP",
	volume = "08",
	pages = "006",
	year = "2012"
}

@article{Moresco:2015cya,
	author = "Moresco, Michele",
	title = "{Raising the bar: new constraints on the Hubble parameter with cosmic chronometers at z {\ensuremath{\sim}} 2}",
	eprint = "1503.01116",
	archivePrefix = "arXiv",
	primaryClass = "astro-ph.CO",
	doi = "10.1093/mnrasl/slv037",
	journal = "Mon. Not. Roy. Astron. Soc.",
	volume = "450",
	number = "1",
	pages = "L16--L20",
	year = "2015"
}

@article{Moresco:2016mzx,
	author = "Moresco, Michele and Pozzetti, Lucia and Cimatti, Andrea and Jimenez, Raul and Maraston, Claudia and Verde, Licia and Thomas, Daniel and Citro, Annalisa and Tojeiro, Rita and Wilkinson, David",
	title = "{A 6{\%} measurement of the Hubble parameter at $z\sim0.45$: direct evidence of the epoch of cosmic re-acceleration}",
	eprint = "1601.01701",
	archivePrefix = "arXiv",
	primaryClass = "astro-ph.CO",
	doi = "10.1088/1475-7516/2016/05/014",
	journal = "JCAP",
	volume = "05",
	pages = "014",
	year = "2016"
}

@article{Brout:2022vxf,
	author = "Brout, Dillon and others",
	title = "{The Pantheon+ Analysis: Cosmological Constraints}",
	eprint = "2202.04077",
	archivePrefix = "arXiv",
	primaryClass = "astro-ph.CO",
	doi = "10.3847/1538-4357/ac8e04",
	journal = "Astrophys. J.",
	volume = "938",
	number = "2",
	pages = "110",
	year = "2022"
}

@article{Goliath:2001af,
	author = "Goliath, M. and Amanullah, R. and Astier, P. and Goobar, A. and Pain, R.",
	title = "{Supernovae and the nature of the dark energy}",
	eprint = "astro-ph/0104009",
	archivePrefix = "arXiv",
	doi = "10.1051/0004-6361:20011398",
	journal = "Astron. Astrophys.",
	volume = "380",
	pages = "6--18",
	year = "2001"
}

@article{DESI:2019jxc,
	author = "Levi, Michael E. and others",
	collaboration = "DESI",
	title = "{The Dark Energy Spectroscopic Instrument (DESI)}",
	eprint = "1907.10688",
	archivePrefix = "arXiv",
	primaryClass = "astro-ph.IM",
	reportNumber = "FERMILAB-PUB-19-434-AE",
	month = "7",
	year = "2019",
	doi = " "
}

@article{eBOSS:2020yzd,
	author = "Alam, Shadab and others",
	collaboration = "eBOSS",
	title = "{Completed SDSS-IV extended Baryon Oscillation Spectroscopic Survey: Cosmological implications from two decades of spectroscopic surveys at the Apache Point Observatory}",
	eprint = "2007.08991",
	archivePrefix = "arXiv",
	primaryClass = "astro-ph.CO",
	doi = "10.1103/PhysRevD.103.083533",
	journal = "Phys. Rev. D",
	volume = "103",
	number = "8",
	pages = "083533",
	year = "2021"
}

@article{Hu:1995en,
	author = "Hu, Wayne and Sugiyama, Naoshi",
	title = "{Small scale cosmological perturbations: An Analytic approach}",
	eprint = "astro-ph/9510117",
	archivePrefix = "arXiv",
	reportNumber = "IASSNS-AST-95-42, CFPA-TH-95-18, UTAP-212",
	doi = "10.1086/177989",
	journal = "Astrophys. J.",
	volume = "471",
	pages = "542--570",
	year = "1996"
}

@article{Arendse:2019hev,
	author = "Arendse, Nikki and others",
	title = "{Cosmic dissonance: are new physics or systematics behind a short sound horizon?}",
	eprint = "1909.07986",
	archivePrefix = "arXiv",
	primaryClass = "astro-ph.CO",
	doi = "10.1051/0004-6361/201936720",
	journal = "Astron. Astrophys.",
	volume = "639",
	pages = "A57",
	year = "2020"
}

@article{Cooke:2018qzw,
	author = "Cooke, Ryan and Fumagalli, Michele",
	title = "{Measurement of the primordial helium abundance from the intergalactic medium}",
	eprint = "1810.06561",
	archivePrefix = "arXiv",
	primaryClass = "astro-ph.CO",
	doi = "10.1038/s41550-018-0584-z",
	journal = "Nature Astron.",
	volume = "2",
	number = "12",
	pages = "957--961",
	year = "2018"
}

@article{Pitrou:2018cgg,
	author = "Pitrou, Cyril and Coc, Alain and Uzan, Jean-Philippe and Vangioni, Elisabeth",
	title = "{Precision big bang nucleosynthesis with improved Helium-4 predictions}",
	eprint = "1801.08023",
	archivePrefix = "arXiv",
	primaryClass = "astro-ph.CO",
	doi = "10.1016/j.physrep.2018.04.005",
	journal = "Phys. Rept.",
	volume = "754",
	pages = "1--66",
	year = "2018"
}

@article{Handley:2015fda,
	author = "Handley, W. J. and Hobson, M. P. and Lasenby, A. N.",
	title = "{PolyChord: nested sampling for cosmology}",
	eprint = "1502.01856",
	archivePrefix = "arXiv",
	primaryClass = "astro-ph.CO",
	doi = "10.1093/mnrasl/slv047",
	journal = "Mon. Not. Roy. Astron. Soc.",
	volume = "450",
	number = "1",
	pages = "L61--L65",
	year = "2015"
}

@article{Handley:2015vkr,
	author = "Handley, W. J. and Hobson, M. P. and Lasenby, A. N.",
	title = "{polychord: next-generation nested sampling}",
	eprint = "1506.00171",
	archivePrefix = "arXiv",
	primaryClass = "astro-ph.IM",
	doi = "10.1093/mnras/stv1911",
	journal = "Mon. Not. Roy. Astron. Soc.",
	volume = "453",
	number = "4",
	pages = "4385--4399",
	year = "2015"
}

@article{Foreman-Mackey:2012any,
	author = "Foreman-Mackey, Daniel and Hogg, David W. and Lang, Dustin and Goodman, Jonathan",
	title = "{emcee: The MCMC Hammer}",
	eprint = "1202.3665",
	archivePrefix = "arXiv",
	primaryClass = "astro-ph.IM",
	doi = "10.1086/670067",
	journal = "Publ. Astron. Soc. Pac.",
	volume = "125",
	pages = "306--312",
	year = "2013"
}

@article{Torrado:2020dgo,
	author = "Torrado, Jesus and Lewis, Antony",
	title = "{Cobaya: Code for Bayesian Analysis of hierarchical physical models}",
	eprint = "2005.05290",
	archivePrefix = "arXiv",
	primaryClass = "astro-ph.IM",
	reportNumber = "TTK-20-15",
	doi = "10.1088/1475-7516/2021/05/057",
	journal = "JCAP",
	volume = "05",
	pages = "057",
	year = "2021"
}

@article{Lewis:2019xzd,
	author = "Lewis, Antony",
	title = "{GetDist: a Python package for analysing Monte Carlo samples}",
	eprint = "1910.13970",
	archivePrefix = "arXiv",
	primaryClass = "astro-ph.IM",
	doi = "10.1088/1475-7516/2025/08/025",
	journal = "JCAP",
	volume = "08",
	pages = "025",
	year = "2025"
}

@article{Akaike:1974vps,
	author = "Akaike, H.",
	title = "{A new look at the statistical model identification}",
	doi = "10.1109/TAC.1974.1100705",
	journal = "IEEE Trans. Automatic Control",
	volume = "19",
	number = "6",
	pages = "716--723",
	year = "1974"
}

@article{wenren2016marginal,
	title={Marginal conceptual predictive statistic for mixed model selection},
	author={Wenren, Cheng and Shang, Junfeng and Pan, Juming},
	journal={Open Journal of Statistics},
	volume={6},
	number={2},
	pages={239--253},
	year={2016},
	doi="10.4236/ojs.2016.62021",
	publisher={Scientific Research Publishing}
}

@article{Trotta:2008qt,
	author = "Trotta, Roberto",
	title = "{Bayes in the sky: Bayesian inference and model selection in cosmology}",
	eprint = "0803.4089",
	archivePrefix = "arXiv",
	primaryClass = "astro-ph",
	doi = "10.1080/00107510802066753",
	journal = "Contemp. Phys.",
	volume = "49",
	pages = "71--104",
	year = "2008"
}

@article{Hussain:2024jdt,
	author = "Hussain, Saddam",
	title = "{Particle production scenario in an algebraically coupled quintessence field with a dark matter fluid}",
	eprint = "2403.10215",
	archivePrefix = "arXiv",
	primaryClass = "gr-qc",
	reportNumber = "2025",
	doi = "10.1016/j.cjph.2025.06.009",
	journal = "Chin. J. Phys.",
	volume = "97",
	pages = "673--695",
	year = "2025"
}

@article{H0LiCOW:2019pvv,
	author = "Wong, Kenneth C. and others",
	collaboration = "H0LiCOW",
	title = "{H0LiCOW {\textendash} XIII. A 2.4 per cent measurement of H0 from lensed quasars: 5.3{\ensuremath{\sigma}} tension between early- and late-Universe probes}",
	eprint = "1907.04869",
	archivePrefix = "arXiv",
	primaryClass = "astro-ph.CO",
	doi = "10.1093/mnras/stz3094",
	journal = "Mon. Not. Roy. Astron. Soc.",
	volume = "498",
	number = "1",
	pages = "1420--1439",
	year = "2020"
}

@article{Hu:2023jqc,
	author = "Hu, Jian-Ping and Wang, Fa-Yin",
	title = "{Hubble Tension: The Evidence of New Physics}",
	eprint = "2302.05709",
	archivePrefix = "arXiv",
	primaryClass = "astro-ph.CO",
	doi = "10.3390/universe9020094",
	journal = "Universe",
	volume = "9",
	number = "2",
	pages = "94",
	year = "2023"
}

@article{H0LiCOW:2016xpx,
	author = "Suyu, S. H. and others",
	collaboration = "H0LiCOW",
	title = "{H0LiCOW {\textendash} I. H0 Lenses in COSMOGRAIL's Wellspring: program overview}",
	eprint = "1607.00017",
	archivePrefix = "arXiv",
	primaryClass = "astro-ph.CO",
	doi = "10.1093/mnras/stx483",
	journal = "Mon. Not. Roy. Astron. Soc.",
	volume = "468",
	number = "3",
	pages = "2590--2604",
	year = "2017"
}

@article{Suyu:2018vqs,
	author = "Suyu, Sherry H. and Chang, Tzu-Ching and Courbin, Fr{\'e}d{\'e}ric and Okumura, Teppei",
	title = "{Cosmological distance indicators}",
	eprint = "1801.07262",
	archivePrefix = "arXiv",
	primaryClass = "astro-ph.CO",
	doi = "10.1007/s11214-018-0524-3",
	journal = "Space Sci. Rev.",
	volume = "214",
	number = "5",
	pages = "91",
	year = "2018"
}

@article{Suyu:2009by,
	author = "Suyu, S. H. and Marshall, P. J. and Auger, M. W. and Hilbert, S. and Blandford, R. D. and Koopmans, L. V. E. and Fassnacht, C. D. and Treu, T.",
	title = "{Dissecting the Gravitational Lens B1608+656. II. Precision Measurements of the Hubble Constant, Spatial Curvature, and the Dark Energy Equation of State}",
	eprint = "0910.2773",
	archivePrefix = "arXiv",
	primaryClass = "astro-ph.CO",
	reportNumber = "SLAC-PUB-13811",
	doi = "10.1088/0004-637X/711/1/201",
	journal = "Astrophys. J.",
	volume = "711",
	pages = "201--221",
	year = "2010"
}

@article{Suyu:2013kha,
	author = "Suyu, S. H. and others",
	title = "{Cosmology from gravitational lens time delays and Planck data}",
	eprint = "1306.4732",
	archivePrefix = "arXiv",
	primaryClass = "astro-ph.CO",
	doi = "10.1088/2041-8205/788/2/L35",
	journal = "Astrophys. J. Lett.",
	volume = "788",
	pages = "L35",
	year = "2014"
}

@article{H0LiCOW:2016qrm,
	author = "Wong, Kenneth C. and others",
	collaboration = "H0LiCOW",
	title = "{H0LiCOW {\textendash} IV. Lens mass model of HE 0435{\ensuremath{-}}1223 and blind measurement of its time-delay distance for cosmology}",
	eprint = "1607.01403",
	archivePrefix = "arXiv",
	primaryClass = "astro-ph.CO",
	doi = "10.1093/mnras/stw3077",
	journal = "Mon. Not. Roy. Astron. Soc.",
	volume = "465",
	number = "4",
	pages = "4895--4913",
	year = "2017"
}

@article{H0LiCOW:2018tyj,
	author = "Birrer, S. and others",
	collaboration = "H0LiCOW",
	title = "{H0LiCOW - IX. Cosmographic analysis of the doubly imaged quasar SDSS 1206+4332 and a new measurement of the Hubble constant}",
	eprint = "1809.01274",
	archivePrefix = "arXiv",
	primaryClass = "astro-ph.CO",
	doi = "10.1093/mnras/stz200",
	journal = "Mon. Not. Roy. Astron. Soc.",
	volume = "484",
	pages = "4726",
	year = "2019"
}

@article{H0LiCOW:2019xdh,
	author = "Chen, Geoff C. -F. and others",
	collaboration = "H0LiCOW",
	title = "{A SHARP view of H0LiCOW: $H_{0}$ from three time-delay gravitational lens systems with adaptive optics imaging}",
	eprint = "1907.02533",
	archivePrefix = "arXiv",
	primaryClass = "astro-ph.CO",
	doi = "10.1093/mnras/stz2547",
	journal = "Mon. Not. Roy. Astron. Soc.",
	volume = "490",
	number = "2",
	pages = "1743--1773",
	year = "2019"
}

@article{Jee:2019hah,
	author = "Jee, Inh and Suyu, Sherry and Komatsu, Eiichiro and Fassnacht, Christopher D. and Hilbert, Stefan and Koopmans, L{\'e}on V. E.",
	title = "{A measurement of the Hubble constant from angular diameter distances to two gravitational lenses}",
	eprint = "1909.06712",
	archivePrefix = "arXiv",
	primaryClass = "astro-ph.CO",
	doi = "10.1126/science.aat7371",
	month = "9",
	year = "2019"
}

@article{H0LiCOW:2019mdu,
	author = "Rusu, Cristian E. and others",
	collaboration = "H0LiCOW",
	title = "{H0LiCOW XII. Lens mass model of WFI2033 {\ensuremath{-}} 4723 and blind measurement of its time-delay distance and H0}",
	eprint = "1905.09338",
	archivePrefix = "arXiv",
	primaryClass = "astro-ph.CO",
	doi = "10.1093/mnras/stz3451",
	journal = "Mon. Not. Roy. Astron. Soc.",
	volume = "498",
	number = "1",
	pages = "1440--1468",
	year = "2020"
}

@article{Kumar:2012gr,
	author = "Kumar, Suresh and Xu, Lixin",
	title = "{Observational constraints on variable equation of state parameters of dark matter and dark energy after Planck}",
	eprint = "1207.5582",
	archivePrefix = "arXiv",
	primaryClass = "gr-qc",
	doi = "10.1016/j.physletb.2014.08.059",
	journal = "Phys. Lett. B",
	volume = "737",
	pages = "244--247",
	year = "2014"
}

@article{Pan:2022qrr,
	author = "Pan, Supriya and Yang, Weiqiang and Di Valentino, Eleonora and Mota, David F. and Silk, Joseph",
	title = "{IWDM: the fate of an interacting non-cold dark matter {\textemdash} vacuum scenario}",
	eprint = "2211.11047",
	archivePrefix = "arXiv",
	primaryClass = "astro-ph.CO",
	doi = "10.1088/1475-7516/2023/07/064",
	journal = "JCAP",
	volume = "07",
	pages = "064",
	year = "2023"
}

@article{Li:2025eqh,
	author = "Li, Tian-Nuo and Zhang, Yi-Min and Yao, Yan-Hong and Wu, Peng-Ju and Zhang, Jing-Fei and Zhang, Xin",
	title = "{Is non-zero equation of state of dark matter favored by DESI DR2?}",
	eprint = "2506.09819",
	archivePrefix = "arXiv",
	primaryClass = "astro-ph.CO",
	month = "6",
	year = "2025",
	doi =""
}

@article{Wang_2024,
	doi = {10.1088/1361-6633/ad2527},
	url = {https://dx.doi.org/10.1088/1361-6633/ad2527},
	year = {2024},
	month = {feb},
	publisher = {IOP Publishing},
	volume = {87},
	number = {3},
	pages = {036901},
	author = {Wang, B and Abdalla, E and Atrio-Barandela, F and Pavón, D},
	title = {Further understanding the interaction between dark energy and dark matter: current status and future directions},
	journal = {Reports on Progress in Physics},
	eprint = "2402.00819",
	archivePrefix = "arXiv",
}

@article{Huang:2025som,
	author = "Huang, Lu and Cai, Rong-Gen and Wang, Shao-Jiang",
	title = "{The DESI DR1/DR2 evidence for dynamical dark energy is biased by low-redshift supernovae}",
	eprint = "2502.04212",
	archivePrefix = "arXiv",
	primaryClass = "astro-ph.CO",
	doi = "10.1007/s11433-025-2754-5",
	journal = "Sci. China Phys. Mech. Astron.",
	volume = "68",
	number = "10",
	pages = "100413",
	year = "2025"
}

@article{Li:2024qso,
	author = "Li, Tian-Nuo and Wu, Peng-Ju and Du, Guo-Hong and Jin, Shang-Jie and Li, Hai-Li and Zhang, Jing-Fei and Zhang, Xin",
	title = "{Constraints on Interacting Dark Energy Models from the DESI Baryon Acoustic Oscillation and DES Supernovae Data}",
	eprint = "2407.14934",
	archivePrefix = "arXiv",
	primaryClass = "astro-ph.CO",
	doi = "10.3847/1538-4357/ad87f0",
	journal = "Astrophys. J.",
	volume = "976",
	number = "1",
	pages = "1",
	year = "2024"
}
	
\end{document}